\documentclass[10pt,journal,compsoc]{IEEEtran}                
\pdfoutput=1
\ifCLASSINFOpdf
  \usepackage[pdftex]{graphicx}
\else
\fi

\usepackage{hyperref}
\usepackage{url}

\hypersetup{
    colorlinks=true,
    citecolor=black,
    linkcolor=black,
    filecolor=black,      
    urlcolor=black,
    }

\newcommand{\tvcg}[1]{{\color{black}{#1}}}
\newcommand{\tvcgminor}[1]{{\color{black}{#1}}}

\begin{document}

\title{Deep Colormap Extraction from Visualizations}
\author{
Lin-Ping~Yuan,~Wei~Zeng,~\IEEEmembership{Member, IEEE},~Siwei Fu,~\IEEEmembership{Member, IEEE},~Zhiliang Zeng,\\Haotian Li,~Chi-Wing~Fu,~\IEEEmembership{Member, IEEE}, and~Huamin Qu,~\IEEEmembership{Member, IEEE}
\IEEEcompsocitemizethanks{
\IEEEcompsocthanksitem
L.-P. Yuan, H. Li, and H. Qu are with the Hong Kong University of Science and Technology. e-mail: \{lyuanaa, haotian.li\}@connect.ust.hk and huamin@cse.ust.hk.

\IEEEcompsocthanksitem
W. Zeng (corresponding author) is with Shenzhen Institutes of Advanced Technology, Chinese Academy of Sciences. e-mail: wei.zeng@siat.ac.cn.
\IEEEcompsocthanksitem
S. Fu is with Zhejiang Lab. 
e-mail: fusiwei339@gmail.com.

\IEEEcompsocthanksitem
Z. Zeng, and C.-W. Fu are with the Chinese University of Hong Kong. e-mail: \{zlzeng, cwfu\}@cse.cuhk.edu.hk.
}
}

\markboth{IEEE TRANSACTIONS ON VISUALIZATION AND COMPUTER GRAPHICS}{Yuan \MakeLowercase{\textit{et al.}}: Deep Colormap Extraction from Visualizations}

\IEEEtitleabstractindextext{
\begin{abstract}
This work presents a new approach based on deep learning to automatically extract colormaps from visualizations.
After summarizing colors in an input visualization image as a Lab color histogram, we pass the histogram to a pre-trained deep neural network, which learns to predict the colormap that produces the visualization.
To train the network, we create a new dataset of $\sim$64K visualizations that cover a wide variety of data distributions, chart types, and colormaps.
The network adopts an atrous spatial pyramid pooling module to capture color features \tvcg{at multiple scales} in the input color histograms.
We \tvcg{then classify the predicted colormap as discrete or continuous}, and refine the predicted colormap based on its color histogram.
\tvcg{Quantitative comparisons to existing methods show the superior performance of our approach on both synthetic and real-world visualizations}.
We further demonstrate the utility of our method with two use cases, \emph{i.e.}, color transfer and color remapping.
\end{abstract}

\begin{IEEEkeywords}
Color extraction, information visualization, deep learning, color histogram
\end{IEEEkeywords}
}

\maketitle
\IEEEdisplaynontitleabstractindextext
\IEEEpeerreviewmaketitle

\IEEEraisesectionheading{
\section{Introduction}}

\IEEEPARstart{C}{olor} is a fundamental visual means to convey information and knowledge in data.
Colormaps are used to characterize the mapping from data domain to colors in visualizations~\cite{card_1999_readings, tominski_2008_task}.
For instance, colors can help depict the attributes in geographical data such as income and dominant commercial sectors~\cite{colorbrewer_2003}.
However, designing effective colormaps has always been challenging, since 
many factors, such as data types, tasks, goals, and users~\cite{tominski_2008_task, silva_2011_using, munzner_2014_visualization}, need to be taken into account simultaneously.

Many color usage guidelines and optimization methods have been proposed to facilitate color design; see~\cite{silva_2011_using, munzner_2014_visualization, liang_2016_survey} for systematic reviews.
Early studies suggested various standards, e.g., a color design should consider the color properties such as uniformity~\cite{robertson_1986_generation} and sequence~\cite{ware_1988_color}, as well as human factors such as perceptual discrimination~\cite{christopher_1996_choosing}.
These guidelines later became fundamentals for \emph{rule-based} colormap selection~\cite{liang_2016_survey}.
Besides these factors, data and tasks should also be carefully considered in the color design process~\cite{liang_2016_survey}, such as for tree-structure data~\cite{tennekes_2014_tree} and categorical data~\cite{gramazio_2017_colorgorical}, to improve the semantic reasoning~\cite{setlur_2016_linguistic, lin_2013_selecting}.

The need for colormap design is immense.
Many colormaps proposed by the above works have been integrated into existing visualization tools~\cite{liang_2016_survey}.
Nevertheless, while inappropriate color ranges could hinder the exploration and analysis of features in data~\cite{liang_2016_survey}, problematic color assignments, which disregard human perception, e.g., rainbow colormaps~\cite{borland_2007_rainbow}, are still ubiquitous.
One promising approach to overcome the issues is to re-style the visualizations~\cite{jung_2017_chartsense, poco_2017_reverse}.
To achieve this, however, requires identification of the color usages in the given visualizations.
Poco et al.~\cite{poco_2018_extracting} developed a semi-automatic approach to extract colormaps from visualization images.
The results benefit many usage scenarios, e.g., to optimize poorly-designed colormaps and to enable interactive overlays to improve readability.
However, the method relies on detecting organized color legends in the visualizations, which may not be available, particularly for those found from the Internet~\cite{yoo_2015_color}.
Thus, some works suggested to directly extract colors from visualizations, e.g., 
by finding a long path of smoothly-varying colors in Lab color space for recovering continuous colormaps~\cite{yoo_2015_color}, or by adapting \textit{k}-means clustering to find discrete colormaps~\cite{chang_2015_palette,zhang_2017_palette}.

However, these methods are heuristic-based and rely on cumbersome predefined parameters, which are prone to fail for general visualizations.
In this work, we present a learning-based approach to automatically extract colormaps from visualizations via a deep neural network.
Instead of adapting the parameters manually, we feed a neural network with a vast amount of visualization images, and let the network learn to model an optimal mapping from the given visualizations to the original colormaps.

An immediate challenge to the approach is the lack of suitable training data.
We hereby prepare a dataset by synthesizing diverse visualization images, with careful consideration of data distributions, chart types, and colormaps (Sec.~\ref{ssec:data}).
Also, we find that directly mapping from visualization images to colormaps is variant to geometric transformations such as rotation.
To overcome the limitation, we first summarize an input visualization image as a color histogram in Lab color space (Sec.~\ref{sssec:histogram}), then adopt an image-translation convolutional neural network (CNN) to produce a fixed-size color legend as an intermediate result (Sec.~\ref{ssec:cnn}).
\tvcg{Specifically, the network employs an atrous spatial pyramid pooling (ASPP) module~\cite{chen_2018_deeplab} that adjusts the filter's field-of-view, being able to capture color features at multiple scales in the input color histogram.}
We further refine the result using DBSCAN or Laplacian eigenmaps to generate the final discrete or continuous colormap (Sec.~\ref{sssec:fine-tune}), respectively.
\tvcg{We conduct quantitative comparisons to existing methods on an evaluation dataset comprised of synthetic and real-world visualizations.
The experimental results demonstrate superior performances of our approach and prove its effectiveness and robustness (Sec.~\ref{sec:evaluation}).}
We further show two applications of our approach: (i) color transfer and (ii) color remapping (Sec.~\ref{sec:app}).
The synthetic visualization image corpus and deep neural network code are available at\tvcgminor{~\url{https://bit.ly/3bjMdyV}}.
\section{Related Work}
\label{sec:related_work}
\textbf{Color design for visualization}.
Designing effective colormaps for visualizations requires both experience and a good sense of aesthetics.
Many research efforts have been devoted to the problem, of which systematic reviews are presented in~\cite{silva_2011_using, munzner_2014_visualization, liang_2016_survey}.
Among these studies, ColorBrewer~\cite{colorbrewer_2003} provides widely-used colormaps. 
While the proposed colormaps have greatly improved the color design in maps and many other applications, the color mapping might miss interesting features in data due to inappropriate color ranges~\cite{liang_2016_survey}.
This promotes the considerations of data and tasks when designing colormaps.
For instance, Lee et al.~\cite{lee_2013_perceptually} showed that the conventional ColorBrewer palette could produce limited visibility of categorical differences in maps, so they improved the color assignment by optimizing the class visibility when presenting coherent categorical structures.
Other factors such as contrast~\cite{mittelstadt_2014_methods} and visual separability~\cite{wang_2019_optimizing} have also been studied to maximize data visualization efficiency.
Furthermore, associating colors with resonant semantics helps discriminate among the data values~\cite{lin_2013_selecting, setlur_2016_linguistic}.

Color design and optimization are out of the scope of this work.
Instead, we focus on exemplar-based visualization design, which has been successfully demonstrated in applications such as image colorization~\cite{revital_2005_colorization, yuan_2021_infocolorizer}, multiple-view visualization design~\cite{chen_2021_composition}, and graph layout tuning~\cite{pan_2021_exemplar, tang_2020_plotthread}.
We explore reverse-engineering approaches to extract colormaps from given visualizations, and also develop two applications that are built upon our method:
(i) transferring color design from source to target visualizations~\cite{welsh_2002_transfer, revital_2005_colorization}, and
(ii) remapping color coding coherently with the underlying data distribution~\cite{tominski_2008_task}.

\vspace{2mm}
\noindent
\textbf{Color extraction from images}.
Extracting colors from images is a basic operation in many applications, e.g., image recoloring, decolorization, etc.
Irony et al.~\cite{revital_2005_colorization} identified the colors in a segmented image sample, and applied the colors to colorize grayscale images.
A typical approach here is to cluster the image colors in the RGB color space then identify the most prominent ones~\cite{chang_2015_palette, zhang_2017_palette}.
Another approach forms a convex hull over the image colors in the color space, and applies geometric processing techniques to extract the primary colors~\cite{tan_2016_decomposing, tan_2018_efficient}.
Both approaches share a common step of summarizing the colors in a certain color space to improve the robustness for images of arbitrary size.

These works, however, target mostly natural images rather than visualizations.
Hence, color design principles for visualizations are not considered, e.g., color ordering for representing ordinal data.
For instance, Chang et al.~\cite{chang_2015_palette} employed a simple policy of sorting all the colors according to the luminance.
In contrast, many existing colormaps order the colors based on their hues.
Yoo et al.~\cite{yoo_2015_color} addressed the ordering problem by fitting a long curve in the CIELab color space to smoothly pass through the colors in a given visualization.
Unfortunately, the method only works for continuous colormaps, {\em not\/} applicable to discrete ones.
Even for continuous colormaps, the method's performance drops when the underlying data distribution is uneven (see evaluations in Sec.~\ref{sec:evaluation}).
Recently, Poco et al.~\cite{poco_2018_extracting} developed a semi-automatic approach to identify color legends in visualizations, classify the legends as discrete or continuous colormaps, and then apply image segmentation techniques to extract the colors from the legends.
Although the method can handle both continuous and discrete colormaps, it assumes the input visualization contains an explicit color legend.
Yoo et al.~\cite{yoo_2015_color} found that only 29\% of 611 web visualization images have a proper color legend.
Image queries on Google search engine with keywords `data visualization', and `information visualization' also show similar results; see Supplementary Fig. 1 for examples.

In summary, prior heuristic-based approaches suffer from tedious parameter settings, such as different numbers of discrete colors~\cite{chang_2015_palette}, and color order recovering~\cite{yoo_2015_color}. 
To overcome the limitations, we develop a learning-based approach, 
 where we prepare a dataset with careful considerations on the data distributions, chart types, and colormaps to supervise the network training.

\vspace{2mm}
\noindent
\textbf{Deep learning for data visualization}.
Advancements of deep learning in various research areas, particularly in computer vision and image understanding, have inspired the visualization community to address various problems; see~\cite{wang2020applying, wu2021survey} for recent surveys.
For example, Jung et al.~\cite{jung_2017_chartsense} trained a deep neural network to classify chart types.
Chen et al.~\cite{chen_2020_lassonet} improved the effectiveness and robustness in selection of 3D point clouds using deep learning.
\tvcgminor{Zhao et al.~\cite{zhao_2020_chartseer} recommended charts for exploratory visual analysis using an interactive system coupled with machine intelligence.}
Some researches further modeled perception-related problems using deep learning. 
\tvcgminor{Wu et al.~\cite{wu2021learning} adopted a Siamese neural network to assess chart layout qualities from pairwise comparison data,} while Haehn et al.~\cite{haehn_2019_evaluating} evaluated various neural networks in analyzing graphical elements.
Besides, some other recent works aim for understanding and diagnosing the learning process in deep neural networks through visual analytics~\cite{Liu2017, Hohman2018}.

Adding on to this line of deep learning for visualization, we develop and train the first deep neural network for colormap extraction from visualization images.
Compared with existing heuristic-based colormap extraction methods, our deep model shows better generalizability and can robustly handle more variety of information visualization images and colormaps.
\section{Overview}
\label{sec:background}

This section clarifies the scope of the work (Sec.~\ref{ssec:scope}), summarizes the problem statement (Sec.~\ref{ssec:problem}), and overviews our method (Sec.~\ref{ssec:system}).

\subsection{Method Scope}
\label{ssec:scope}

Recovering visual encodings from visualizations has recently gained much attention in the visualization community~\cite{poco_2017_reverse}.
Some studies have exploited structures in interactive D3 visualizations in SVGs~\cite{harper_2018_converting, hoque_2019_searching}.
Instead, we constrain the scope of this work to static bitmap images, which are the most commonly available format for charts, and accurate extraction remains challenging~\cite{harper_2018_converting}.
More specifically, we focus on information visualization, or abstract two-dimensional data visualization, e.g., bar charts, maps, and scatter plots, etc.
Visualizations for scientific data (e.g., particles, streamlines, and volumes) usually depict depth or motion with transparency, which, however, may distract the color saturation and luminance~\cite{munzner_2014_visualization}.
We leave it as a future work to extract colormaps from scientific visualizations.

In the early stage of this work, we conducted literature surveys on color usage in information visualization.
Based on assorted systematic reviews~\cite{silva_2011_using, munzner_2014_visualization, liang_2016_survey}, we further make careful specifications on the scope of this work:

\begin{itemize}

\item
\textit{1D}~\emph{v.s.}~\textit{2D colormaps}.
A variety of 2D colormaps have been proposed for visualizing bivariate or multivariate data.
However, they were criticized due to the difficulty in data interpretation, especially when both attributes have multiple levels~\cite{bernard_2015_survey, munzner_2014_visualization}.
In this work, we only consider 1D colormaps, which are the most commonly-used colormaps in the production of information visualization~\cite{liang_2016_survey}.

\item
\textit{Blending} or \textit{interpolation}.
When visualizing multivariate data, some frequently-used techniques are color blending or interpolation~\cite{cheng_2019_colormap}.
Both techniques will generate nonlinear mappings of colors, which may hinder the deep neural network to learn the underlying colormaps.

\item
\textit{Background \& text colors}.
The presence of dissimilar colors near one another can greatly affect the color perception~\cite{mittelstadt_2015_methdos}.
To create effective colormap extraction, background and text colors should be removed before processing.

\item
\textit{Linear}~\emph{v.s.}~\textit{non-linear \tvcg{color scales}}.
There are some perceptually non-linear \tvcg{color scales that allot more colors to data of interest and fewer colors to all other data}, to cope with \tvcg{skewed} data distributions~\cite{tominski_2008_task}.
Such non-linear \tvcg{interpolation between data and colors} can cause ambiguity with \tvcg{commonly available linear color scales}.
Hence, this work regards all \tvcg{color scales} behind input visualizations as linear.
That is, we do not extract how data values map to colors.
\end{itemize}

\subsection{Problem Statement}
\label{ssec:problem}

We classify colormaps as {\em discrete\/} or {\em continuous\/}.
A discrete colormap contains only a small set of colors, while a continuous colormap is essentially a curve in a color space.
We model both discrete and continuous colormaps as an ordered list of $m$ color values, denoted as $C := \{c^i\}_{i=1}^{m}$, where $i$ indicates the $i$-th color in $C$ and $m$ is small, typically less than 10 for discrete colormaps~\cite{ware_2010_visual}, and we fix it as 256 for continuous colormaps.
The setting is pragmatic, as continuous colormaps are usually presented as a 1D image with smoothly-varying colors.

In information visualizations, colormaps are employed to map data values to colors.
We treat the original colormap as the ground truth $C_{gt} := \{c^i_{gt}\}_{i=1}^{m}$ 
and aim to develop a new algorithm to automatically extract colormap $C_{out} := \{c^j_{out}\}_{j=1}^{n}$ that is similar to $C_{gt}$.
Notice that $C_{out}$ may not have the same number of colors as $C_{gt}$.
We consider the following requirements for such an algorithm:

\begin{itemize}

\item
\textit{Effectiveness.}
The algorithm should generate $C_{out}$ in color space $S$, and $C_{out}$ should be similar to $C_{gt}$, meaning that
%
\begin{equation}
\label{eq:min_Cout}
\min_{C_{out} \in S} ~ D(C_{gt}, C_{out}) \ ,
\end{equation}
where \textit{D} is a distance function.
Specifically, we should preserve the \textit{ordering} of colors in each colormap, which is essential for ordinal data.
To this end, we employ \tvcg{dynamic time warping (DTW) that is a popular technique for comparing sequences~\cite{giorgino2009computing}}, to model $D$ in Eq.~\ref{eq:min_Cout}; see Sec.~\ref{ssec:quan_comp} for details.

\item
\textit{Robustness.}
The method should
(i) be applicable to both discrete and continuous colormaps;
(ii) handle visualizations with or without given color legends;
(iii) deal with various types of information visualizations, such as bar charts, pie charts, and line plots; and
(iv) be invariant to geometric transformation such as rotation.

\end{itemize}

\begin{figure}[t]
	\centering
	\includegraphics[width=1\linewidth]{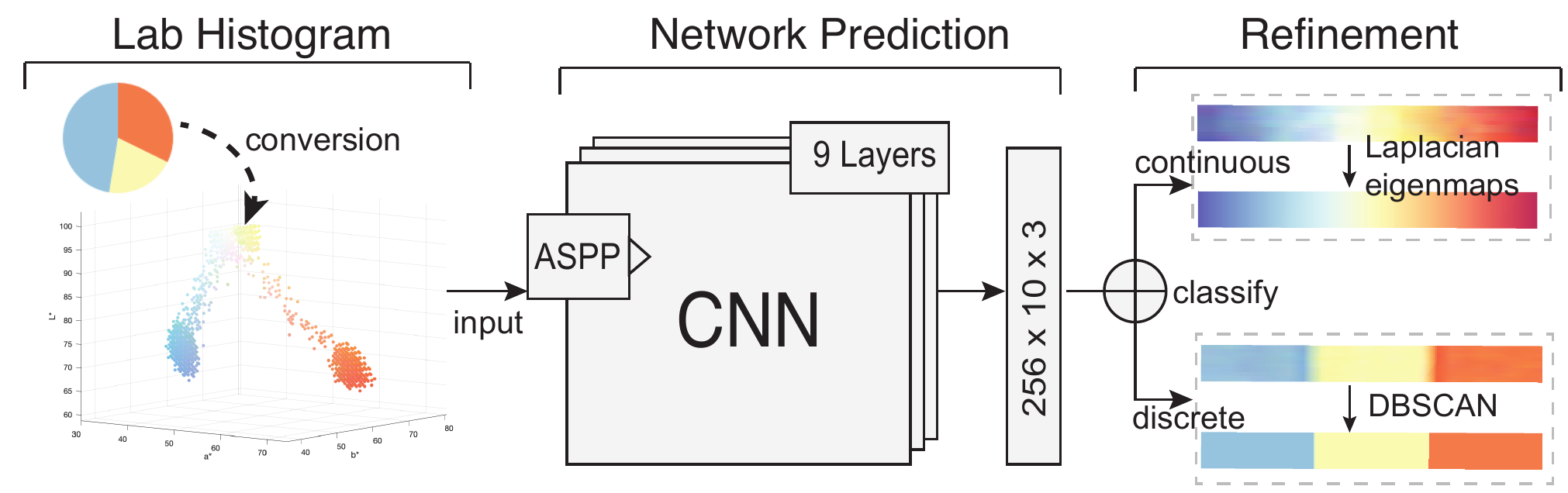}
	\caption{Our approach consists of three stages: (i) histogram conversion, (ii) network prediction, and (iii) refinement.
	}
	 \vspace{-4mm}
	\label{fig:overview}
\end{figure}

\subsection{Technique at Large}
\label{ssec:system}

Extracting colormaps from given visualizations is a reverse-engineering process.
In contrast to prior heuristic-based methods, we approach the problem from a new perspective, by regarding the process as a mapping problem, \emph{i.e.}, to map a visualization to the associated colormap.
We hereby adopt and train a deep neural network model, and drive it to learn with carefully-prepared example visualizations paired with the underlying colormaps.

As shown in Fig.~\ref{fig:overview}, our approach has three major stages:

\begin{enumerate}

\item
\textit{Histogram conversion} (Sec.~\ref{sssec:histogram}).
Preliminary experiments reveal that directly mapping from visualization images to colormaps is highly variant against rotation (Fig.~\ref{fig:rotation}).
To facilitate the network to learn the color features instead of other visual primitives such as shapes and locations, we first summarize the input visualization image into a color histogram in the Lab color space.
This conversion can effectively reduce the number of variants in the network input, and ensure the network is rotation invariant.

\item
\textit{Network prediction} (Sec.~\ref{ssec:cnn}).
The network takes a color histogram as input and generates a fixed-size image as its output.
Nevertheless, there exist several color-related issues, e.g., null features in the input histograms.
We make adaptations to the baseline \tvcg{ResNet18 network~\cite{he2016deep}}, 
including the incorporation of the ASPP module~\cite{chen_2018_deeplab}, to tackle the challenges.

\item
\textit{Refinement} (Sec.~\ref{sssec:fine-tune}).
Lastly, we refine the output network prediction and produce an ordered list of color values.
The refinement is a two-fold process.
First, we utilize a binary classification rule to determine if the prediction is a continuous or discrete colormap.
Second, we apply Laplacian eigenmaps or DBSCAN to extract the output continuous or discrete colormaps, respectively.

\end{enumerate}

\section{Deep Colormap Extraction}
\label{sec:deep}

To train a deep neural network, we first prepare training data that covers a wide variety of information visualizations, each paired with a corresponding colormap (Sec.~\ref{ssec:data}). 
Next, we present the technical details in our deep colormap extraction method (Sec.~\ref{ssec:methods}).

\subsection{Data Generation}
\label{ssec:data}

While some visualization corpora are available (e.g.,~\cite{savva_2011_revision, borkin_2013_what, jung_2017_chartsense, poco_2018_extracting}), they do not provide the paired colormaps.
It would require tremendous labor work to manually reconstruct the underlying colormap for each given visualization.
Motivated by the use of image synthesis to generate training datasets~\cite{ma_2018_scatternet, haehn_2019_evaluating},
we develop an automatic method to synthesize information visualizations with corresponding colormaps.
Thanks to the wide usage of visualization toolkits like D3 in real-world applications, the synthetic data can mimic real-world visualizations~\cite{poco_2017_reverse}.

We next elaborate on the considerations for maximizing the coverage of the sample information visualizations.
Then, we further present the technical details of synthesizing a new dataset, which contains $\sim$64K visualizations with appropriate combinations of realistic data distributions, chart types, and colormaps.

\vspace{2mm}
\noindent
\textbf{Design considerations}.
Visualization creation can be viewed as an exploratory process in the \textit{control} space (which includes the visualization styles, layout, colormaps, etc.) until satisfactory \textit{visualization} results are produced for the input \textit{data}~\cite{chen_2009_data}.
That is to say, a visualization depends both on the input data and the control parameters.
Specifically, we diversify the control parameters in two aspects: (i) chart types produce variants in the geometric space, e.g., lines~\emph{vs.}~areas, and (ii) colormaps complement the geometric diversity in the color space.

\begin{figure}[tb]
    \centering
    \includegraphics[width=0.9\linewidth]{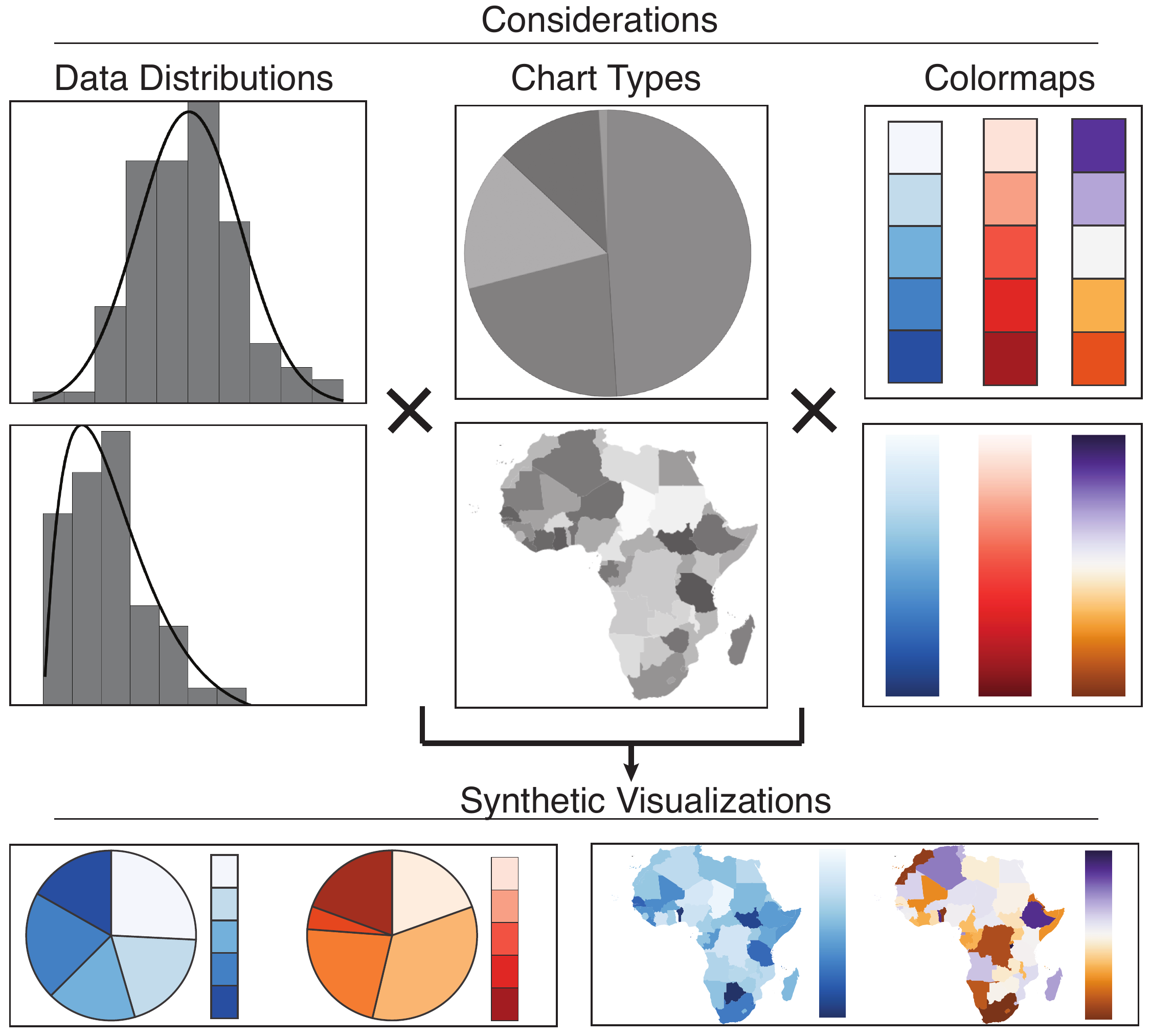}
    \vspace{-2mm}
    \caption{
    We carefully consider various variants in data distributions, chart types, and colormaps (top), when generating the synthetic visualizations with the corresponding colormaps (bottom).
    }
     \vspace{-4mm}
    \label{fig:data_generation}
\end{figure}

Fig.~\ref{fig:data_generation} illustrates how we produce diverse visualizations, for example, through
2 data distributions (normal and beta) $\times$ 2 chart types (pie and choropleth map) $\times$ 6 colormaps (3 discrete and 3 continuous) = 24 visualizations.
Among the combinations, some are non-practical, e.g., pie charts with continuous colormaps.
Nevertheless, even without such combinations, there are over 10 combinations remaining, four of which are presented in Fig.~\ref{fig:data_generation} (bottom). 

\vspace{2mm}
\noindent
\textbf{Implementation details}.
To enrich the dataset coverage, we first create vast variants in the following aspects.
\begin{itemize}

\vspace*{1mm}
\item
\textit{Data distributions}.
To cope with real-world scenarios, we first select 1,041 datasets with attribute number in the range of [1, 10] from the Rdatasets\footnote{\url{https://vincentarelbundock.github.io/Rdatasets/}}, which contains over 1,300 popular datasets used by statisticians and visualization researchers, e.g., the famous \textit{Iris} flower data;
see Supplementary Fig. 2 for more details.
Second, we enrich the dataset with synthetic one- and two-dimensional data.
One-dimensional data follows the common distributions such as normal, beta, and Poisson distributions, while two-dimensional data can either be independent or follow a linear or inverse-linear correlation.
For each distribution type, we create 10 sets of random parameters, yielding a set of data distribution variants.
All these data distributions are generated using SciPy\footnote{\url{https://www.scipy.org}}.

\vspace*{1mm}
\item
\textit{Chart types}.
We first study common charts as identified by prior researches, e.g., Revision~\cite{savva_2011_revision} and ChartSense~\cite{jung_2017_chartsense}.
Among these charts, Venn diagrams and radar charts usually incorporate translucency in colors, which are ignored for now.
We also include other chart types such as heatmaps and choropleth maps, resulting in a total of eight chart types, \emph{i.e.}, \{\emph{line chart}, \emph{pie chart}, \emph{grouped bar chart}, \emph{stacked bar chart}, \emph{scatter plot}, \emph{stream graph}, \emph{heat map}, and \emph{choropleth map}\}.
For each chart type, we further consider different parameter settings, such as line width in line charts, and spacing between bars in bar charts, by referencing to common visualization libraries, including D3 and Vega, and applications such as Tableau and MS Excel.

\vspace*{1mm}
\item
\textit{Colormaps}.
We look for frequently-used colormaps from various visualization libraries, \tvcg{including ColorBrewer~\cite{colorbrewer2}, colorcet~\cite{colorcet}, and D3~\cite{d3_scale_chromatic}.
We constrain the number of colors to be three to ten for discrete colormaps, since such number is already sufficient to cover most visualizations of discrete data; see Supplementary Fig. 2.}
\tvcgminor{Duplicates from different sources are manually removed after approximate comparisons.}
In the end, we collect 236 distinct colormaps, among which 54 are continuous, and the remainings are discrete.
For each colormap, we create an ordered list of Lab colors for final quantitative evaluation, and a 10$\times$256 colormap image derived from the color list to supervise the network training.

\end{itemize}

For each rational combination of these variants, we render one visualization using D3, and save it using a backend server.
In this way, we generate a total of 63,965 visualization images.
All images are in the size of 512 $\times$ 256.

\subsection{Building Deep Colormap Extraction Model}
\label{ssec:methods}

With the synthetic dataset, we can proceed to the next phase of learning the mapping from visualizations to colormaps.
We formulate a new deep colormap extraction model that consists of three main stages: \textit{histogram conversion} (Sec.~\ref{sssec:histogram}), \textit{CNN prediction} (Sec.~\ref{ssec:cnn}), and \textit{refinement} (Sec.~\ref{sssec:fine-tune}).

\subsubsection{Histogram Conversion}
\label{sssec:histogram}

\begin{figure}
\centering
\includegraphics[width=0.995\linewidth]{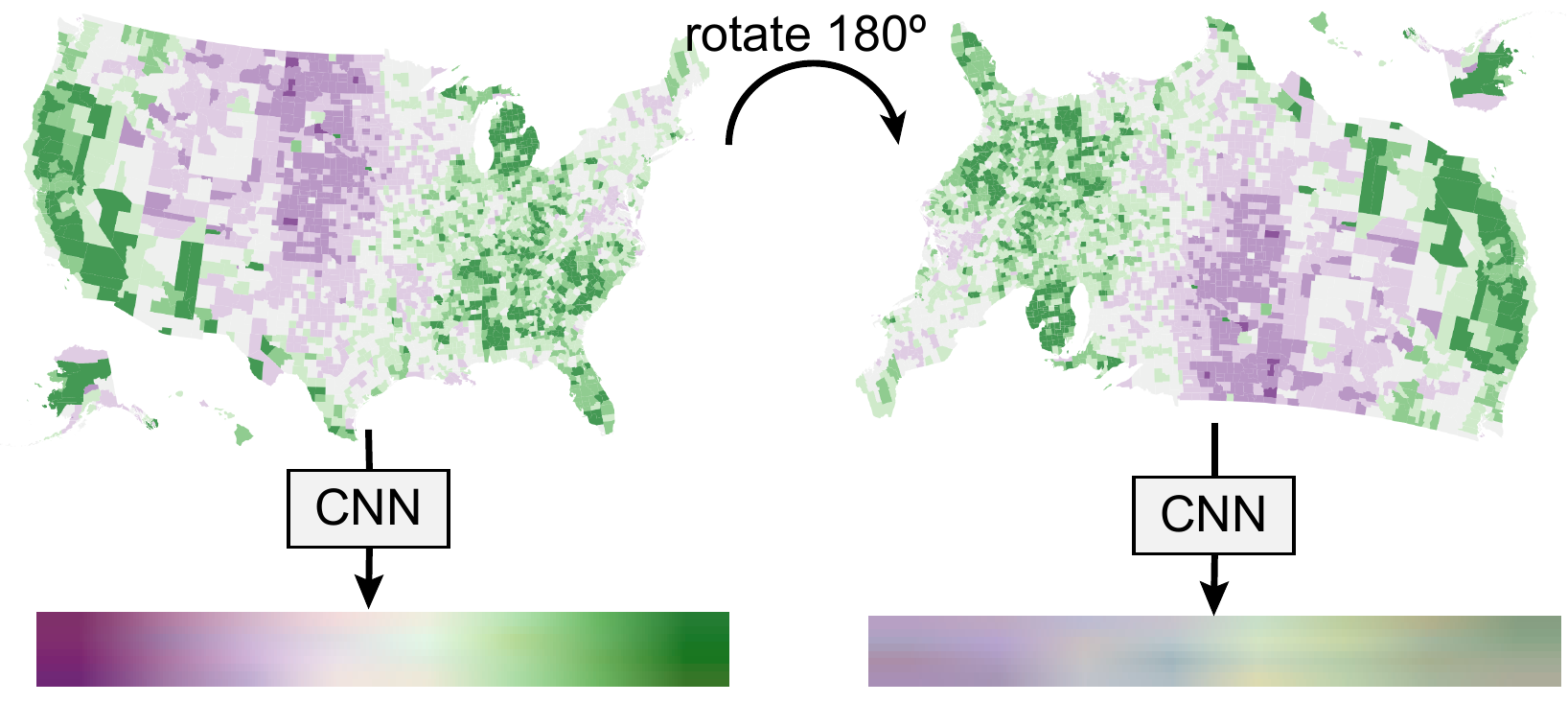}
\vspace{-2mm}
\caption{
A neural network that directly learns colormaps from visualization images is variant to image rotation.
}
\label{fig:rotation}
\vspace{-3mm}
\end{figure}

A vital requirement of neural networks for visualization image analysis is to preserve the geometric invariance, like rotation and scale~\cite{haehn_2019_evaluating}.
We conducted preliminary experiments of training a CNN model to directly map visualization images to colormaps.
However, results showed that the model cannot fulfill the requirement of being rotation invariant, as demonstrated in Fig.~\ref{fig:rotation}.
To eliminate the effects by geometric features, \emph{e.g.}, shapes and positions, we opt to summarize colors in an input visualization image as a 3D histogram in the CIELab color space, then convert the 3D histogram into a 2D map format that is preferable for existing CNN architectures.
Here, the CIELab color space is chosen because of its perceptual uniformity.
The conversion works as follows:

\begin{enumerate}
\vspace{1mm}
\item
\textit{Construct color histograms}. \
We divide the $L$ channel of the CIELab color space into 256 bins, and $a$ and $b$ channels into 128 bins, yielding two $256 \times 128$ color histograms: one for \emph{L-a} and the other for \emph{L-b}.
Here, each histogram entry stores the number of pixels in the input visualization image that has the corresponding \emph{L-a} or \emph{L-b} values.

\vspace{1mm}
\item
\textit{Filter-and-normalize}. \
Next, we filter out the colors for background, text, and grid elements by zero-ing the histogram entries that correspond to these colors.
For the background color, we traverse pixels at the visualization image borders and look for a dominant color that occupies $\geq 80\%$ of all these pixels.
For text and grid element colors, we make a similar assumption as~\cite{poco_2018_extracting} that they are black.
Notice that these operations filter out colors that are rare in common colormaps, even though some colors may seem the same (\emph{e.g.}, dark gray \emph{vs.} black).
Then we normalize all entries in the \emph{L-a} and \emph{L-b} histograms in the range [0,1] by dividing the maximum value.

\vspace{1mm}
\item
\textit{Concatenation}. \
Lastly, we concatenate the two histograms into a single 2D map of size 256$\times$256 as the network input.
Essentially, this is a single-layer 2D map that summarizes the color information in the input visualization for the network to learn to map it to the target colormap.

\end{enumerate}

\begin{figure*}[t]
    \centering
    \includegraphics[width=0.995\linewidth]{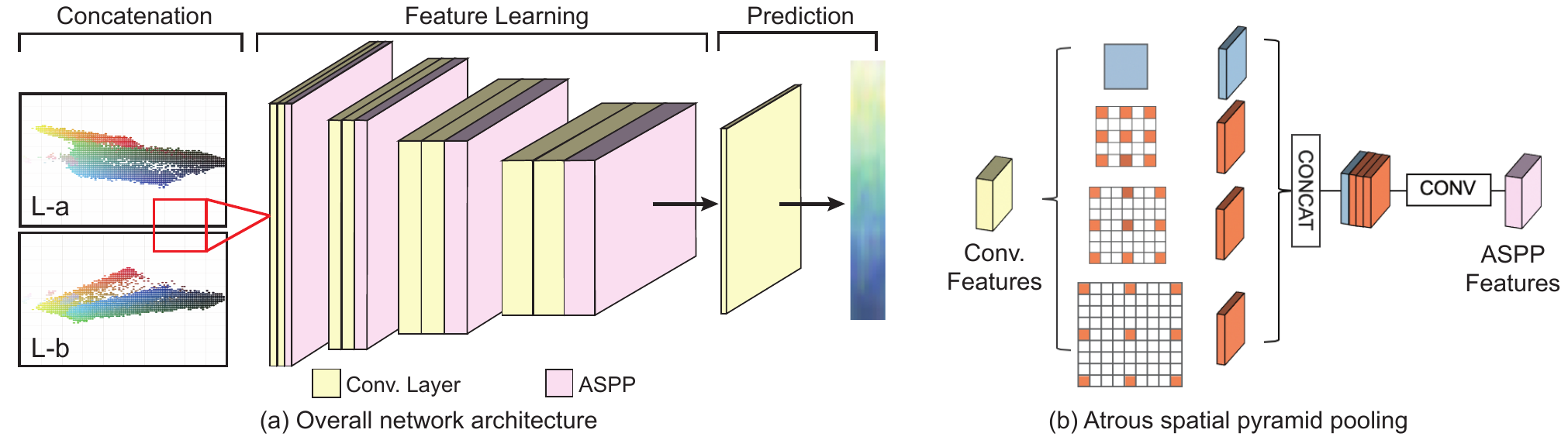}   
    \vspace{-2mm}
    \caption{The architecture of the deep neural network we adopted is presented in (a).
We adopt four stages of convolutional layers (yellow boxes) for feature extraction and feature learning, and make use of 
ASPP modules (pink boxes) after each stage to further distill the features using a pyramid kernel. \ 
After that, to predict the output colormap, we use a 1$\times$1 kernel-size convolutional layer and a bilinear resize to generate an image output, which is then resized to generate the output colormap.
The details of the ASPP module is shown in (b).
}
    \label{fig:architecture}    
     \vspace{-3mm}
\end{figure*}

\subsubsection{CNN Prediction}
\label{ssec:cnn}

\textbf{Network architecture}.
\tvcg{We build the network based on \tvcg{ResNet18~\cite{he2016deep}}.}
The overall architecture is shown in Fig.~\ref{fig:architecture}.
First, we adopt four stages of convolutional layers.
Each stage has two convolutional layers, each activated by a ReLu function and using a fixed 3$\times$3 kernel size.
After each stage, the feature map size is halved and the depth doubles.
Through this hierarchical structure, the network is able to progressively distill the features.
Furthermore, we adopt the ASPP module~\cite{chen_2018_deeplab}, after each stage of convolutional layers.
Since, most values (the given color histogram) in the network input are zeros, the null values hinder the network to learn effectively, and may even cause it to collapse, i.e., poorly converge.
Hence, we replace the fixed-size kernel with a pyramid kernel provided the ASPP module; by using a larger kernel, the network can have a larger receptive field to distill larger-scale (more nonlocal) features.
After such feature learning, we employ a convolutional layer with a 1$\times$1-kernel to further process the extracted features.
Lastly, we generate the output colormap from the last layer through a sigmoid function, and bilinearly resize the output to produce a colormap of size 10$\times$256$\times$3.
By modeling the output as a fixed-size image, our network can effectively handle the conflict of limited neurons \emph{vs.} enormous colors.

\vspace{2mm}
\noindent
\textbf{Loss function.} 
To supervise the network learning, we use an $L_2$ loss to measure the distance between the network output and the ground truth colormap, since both are simply 2D images by nature:
\vspace{-2mm}
\begin{equation}
\label{eq:l2loss}
Loss_{L2}(X,Y) = \frac{1}{n} \sum_{i=1}^{n}|| x_{i} - y_{i} ||^2 \ ,
\end{equation}

\vspace{-1mm}
where $x_i$ and $y_i$ are the colors of the $i$-th pixel on the network output $X$ and the ground truth colormap $Y$, respectively.
The pixel distance is measured as Euclidean distance in Lab color space, where all channels are normalized in the range [0, 1].

\vspace{2mm}
\noindent
\textbf{Implementation details.}
We synthesize 63,965 visualizations in total, where we use 90\% for training and 10\% for testing.
The network is implemented using PyTorch and trained on a workstation with NVIDIA GeForce GTX 1080 Ti GPU.
We employ the Adam optimizer to train the network for 80,000 iterations with a learning rate of 0.0001, and a batch size of 8.
We do not use any batch normalization nor layer normalization.

\begin{figure}
	\centering
	\includegraphics[width=0.995\linewidth]{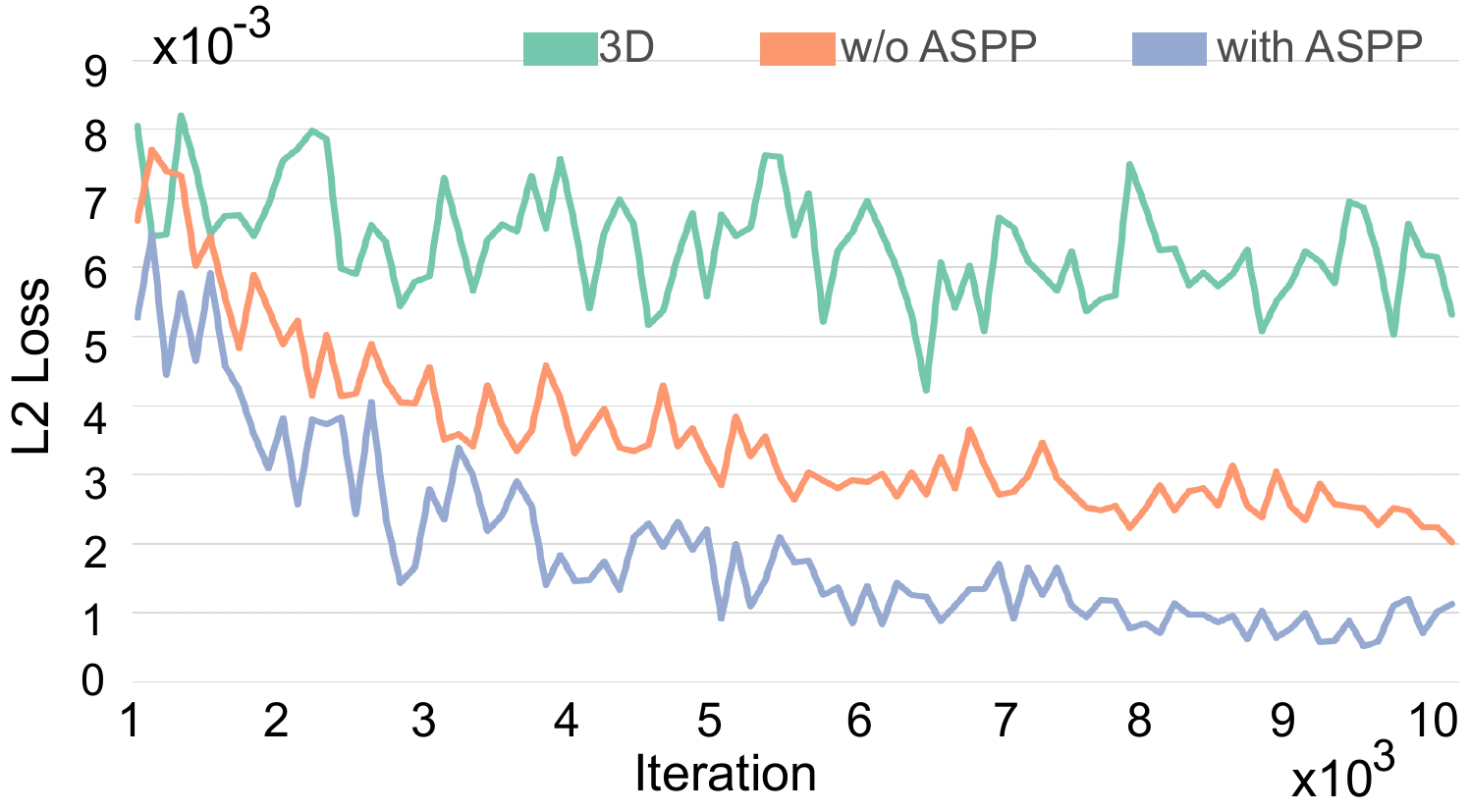}
	\vspace{-3mm}
	\caption{Ablation analysis of comparing network input as 3D tensor (green) \emph{vs.} 2D map concatenation without ASPP module (orange) \emph{vs.} 2D map concatenation with ASPP module (blue). 
	}
	 \vspace{-3mm}
	\label{fig:aspp}	
\end{figure}

\vspace{2mm}
\noindent
\textbf{Ablation analysis}.
We perform an ablation analysis on the model loss over training iterations to evaluate the effectiveness of modeling the Lab histogram as a 2D map concatenation, and the ASPP module.
Fig.~\ref{fig:aspp} presents the comparison results: green curve for network input as 3D tensor without ASPP, orange curve for network input as 2D map concatenation without ASPP, and blue curve for a network with ASPP.
All other hyperparameter settings employed in the networks are the same, including the training data, loss function, and learning rate, etc.
We measure $L_2$ loss using 100 testing cases at every 100 training iterations for the first 10,000 iterations.

First, instead of using a 2D map concatenation, we can feed into the network a 3D tensor that directly represents the Lab histogram.
We notice that the performance here is not comparable with 2D map concatenation.
We suspect this is because general CNNs consume 3D tensor as stacks of 2D feature maps, rather than geometric features in the 3D space of the Lab histograms.
Second, a comparison between the orange and blue curves shows that the network with the ASPP module can converge faster and achieve better result:
the loss is twice in the network without ASPP (2.4$\times10^{-3}$) than with ASPP (1.1$\times10^{-3}$) after iteration 10,000.

\tvcg{Our current implementation of the network can be regarded as a regression function.
Alternatively, we can model the problem as a multi-class classification task that directly categorizes input histograms into classes of colormaps.
We train a classification network with the same input and internal structure of that in Fig.~\ref{fig:architecture}, but change the loss function to a multi-class cross entropy loss with 236 classes of colormaps.
However, for real-world visualization, we find the classification network can easily misclassify discrete and continuous colormaps or predict colormaps with wrong color schemes, and can never recognize unseen colormaps.
Our regression approach is more robust and generalizable.}

More results on ablation analyses are discussed in Supplementary Sec. 3.

\begin{figure}[t]
	\centering
	\includegraphics[width=0.995\linewidth]{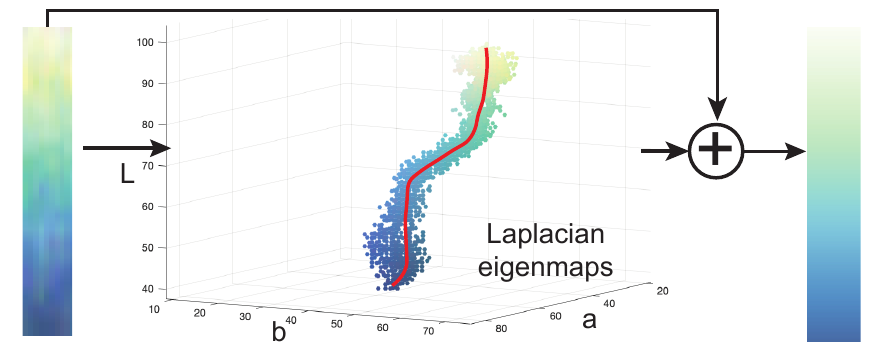}
	\caption{Refinement workflow: CNN prediction (left), Laplacian eigenmaps for continuous colormaps (middle), and final result (right).}
	 \vspace{-3mm}
	\label{fig:fine_tune}	
\end{figure}

\subsubsection{Refinement}
\label{sssec:fine-tune}

Fig.~\ref{fig:fine_tune}(left) presents a network prediction of YlGnBu continuous colormap,
which however is rather noisy.
We refine the prediction to the final result (Fig.~\ref{fig:fine_tune}(right)):

\begin{enumerate}

\item
\textit{Binary classification.}
We check if the prediction colormap is discrete or continuous.
This can be done using a simple binary classification rule: 
we first aggregate histograms of the prediction result in Lab color space and normalize the histograms into the range [0, 1].
Then we filter out all bins with a normalized value smaller than 0.01.
After that, we empirically classify the prediction colormap as discrete or continuous based on the number of remaining histograms.

\vspace{1mm}
\item
\textit{Color extraction}.
Next, we apply DBSCAN and Laplacian eigenmaps to the filtered discrete and continuous colors, respectively.
For DBSCAN, we experimentally find that $minNum = 4$ and $\varepsilon = 0.05$ yield good results for clustering discrete colors.
In each cluster, the color with maximum histogram is chosen as the prominent color.
For Laplacian eigenmaps, we closely follow the procedures of dimension reduction as in ~\cite{yoo_2015_color}.

\vspace{1mm}
\item
\textit{Ordering recovery}.
Both DBSCAN and Laplacian eigenmaps algorithms extract unordered colors.
In the last step, we recover color order by referring to the prediction results that store the ordering information.
To do so, we find all pixels that have the same color as those extracted from the previous step, then measure the average pixel position with respect to prediction results for each extracted color.
Finally, we sort the extracted colors in ascending order of the average pixel position.

\end{enumerate}

\section{Evaluation}
\label{sec:evaluation}

We evaluate the \textit{effectiveness} and \textit{robustness} of our deep color extraction model.
\tvcg{The evaluation is conducted on a testing dataset with both synthetic and real-world visualizations (Sec.~\ref{ssec:eval_data}).}
We present advancements of our approach over existing methods using quantitative metrics (Sec.~\ref{ssec:quan_comp}) and representative examples (Sec.~\ref{ssec:example}).
Last, we analyze probable reasons for the advancements (Sec.~\ref{ssec:analysis}).

\tvcg{
\subsection{Preparation of Testing Dataset}
\label{ssec:eval_data}
We extract 10\% of the synthetic visualizations for evaluation, yielding 6,396 pairs of visualizations and colormaps.
The visualizations cover all the colormaps and chart types we used for training; see Sec.~\ref{ssec:data}. 
We denote these visualizations as \textbf{synthetic testing} in the following.

We also collect additional visualization from the Internet to further evaluate our approach on real-world visualizations.
The collection process consists of two steps:

\begin{itemize}
    \item 
    \textit{Visualization Collection}.
    We first crawl over 1000 visualizations from Google Images searching engine that covers media, websites, and government reports.
    Then, we filter out visualizations out of the work scope described in Sec.~\ref{ssec:scope}, and obtain 229 valid visualizations.
    The resulting visualizations cover both the chart types in the synthetic visualizations and others like radial plots, sunbursts, contours, etc.

    \item
    \textit{Color Labeling}.
    To support quantitative evaluation, we manually label all the collected charts to serve as ground truth.
    All visualizations using continuous colormaps have legends, over 90\% of which are explicitly presented in the charts and the rest can be contained from the source websites.
    For visualizations using discrete colormaps, we retrieve individual colors from explicit legends or the charts themselves, and store the colors in order.
\end{itemize}

We collect in total 229 real-world visualizations, among which 126 visualizations use colormaps in the training dataset (denoted as \textbf{real seen}), whilst colormaps of the remaining 103 visualizations are not in the training dataset (denoted as \textbf{real unseen}).
We conduct experiments on the testing dataset, and report the results by categories of \textbf{synthetic testing, real seen, and real unseen}, respectively.
}

\subsection{Quantitative Comparison}
\label{ssec:quan_comp}

\textbf{Quantitative metric:}
We model both continuous and discrete colormaps as an ordered list of colors $C := \{c^i\}_{i=1}^{m}$ (Sec.~\ref{ssec:problem}).
To account for color ordering, we employ \tvcg{dynamic time warping (DTW)} to calculate the distance (denoted as $D_{dtw}$) between the ground truth colormap $C_{gt} := \{c^i_{gt}\}_{i=1}^{m}$ and the output colormap $C_{out} := \{c^j_{out}\}_{j=1}^{n}$.
\tvcg{DTW is a commonly-used similarity measuring algorithm for two series, since the algorithm is insensitive to local compression and stretches, and can optimally deform one of the two input series onto the other~\cite{giorgino2009computing}.}

To compute $D_{dtw}$, we first generate a coupling between $C_{gt}$ and $C_{out}$, which is a set $L$ of pairings $\{\delta_k\}_{k=1}^{K}$, where $\delta_k = (c_{gt}^k, c_{out}^k) \in [m]\times[n]$.
We can compute:
\begin{equation}
\label{eq:dtw}
    D_{dtw}(C_{gt}, C_{out}) = \min_L \sum_{\delta_k\in L}\, d(c_{gt}^k, c_{out}^k),
\end{equation}
\noindent
where $d(c_{gt}^k, c_{out}^k)$ is the Euclidean distance of $c_{gt}^k$ and $c_{out}^k$ in the normalized Lab color space (all channels are normalized to [0, 1]).
\tvcg{Smaller $D_{dtw}$ indicates better performance.
We adopt the \textit{dtw} package~\cite{giorgino2009computing} to compute $D_{dtw}$.}

\vspace{1mm}
\noindent
\tvcg{\textbf{Baseline techniques:}}
We compare the performance of our approach with two existing methods for colormap extraction, \emph{i.e.}, palette-based~\cite{chang_2015_palette} and sequence-preserving~\cite{yoo_2015_color}.
We omit the legend-based approach~\cite{poco_2018_extracting} \tvcg{on synthetic visualizations, as the method requires explicit color legends in the image space that are not available in our synthetic visualizations.} 
On the other hand, both palette-based~\cite{chang_2015_palette} and sequence-preserving~\cite{yoo_2015_color} methods, as well as ours, process visualization images in the color space.
\tvcg{Nevertheless, a quantitative comparison with~\cite{poco_2018_extracting} on real-world visualizations with explicit legends can be found in the Supplementary Sec. 4.2.}
For a fair comparison, we only evaluate the palette-based method~\cite{chang_2015_palette} on discrete colormaps, and the sequence-preserving method~\cite{yoo_2015_color} on continuous colormaps.
Specifically, we re-implemented sequence-preserving method~\cite{yoo_2015_color} in MATLAB, and employed an online re-implementation \footnote{\url{https://github.com/b-z/photo_recoloring}} of the palette-based method~\cite{chang_2015_palette}.

\begin{figure}
	\centering
	\includegraphics[width=1\linewidth]{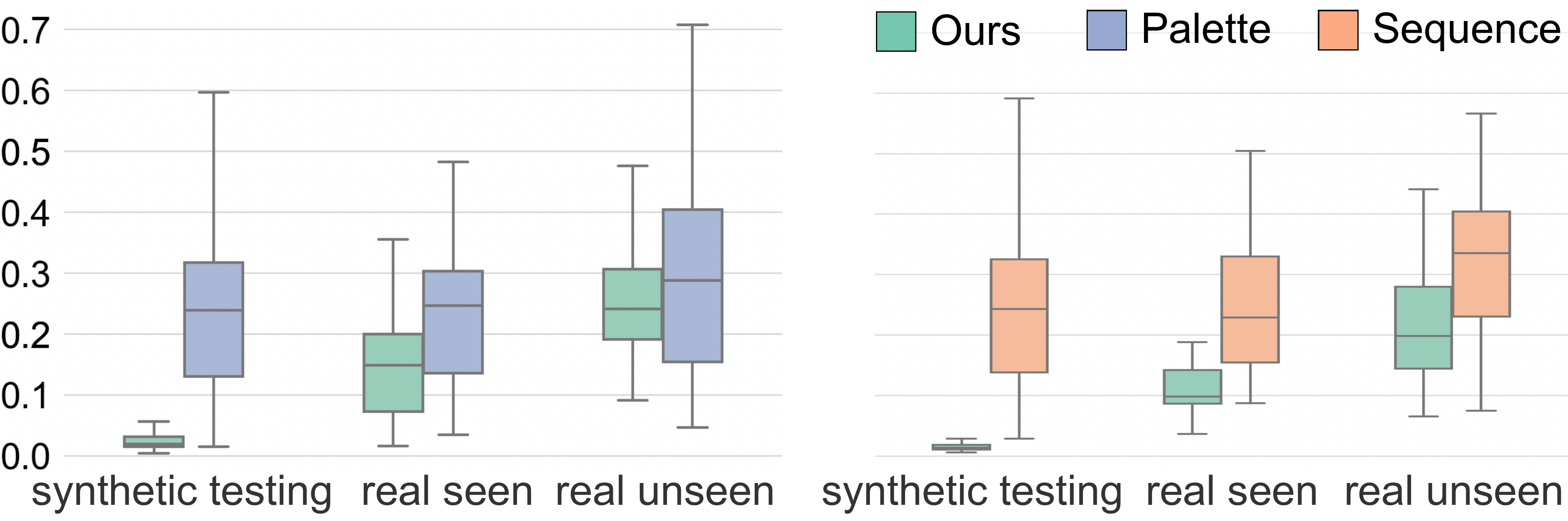}
	\vspace{-5mm}
	\caption{\tvcg{Comparing $D_{dtw}$ on synthetic testing, real seen, and real unseen visualizations. 
	As indicated by the smaller $D_{dtw}$ values,
    our method achieves better performances on all three categories of visualizations than the baseline techniques (Palette~\cite{chang_2015_palette} for discrete colormaps (left), and Sequence~\cite{yoo_2015_color} for continuous colormaps (right)), especially for the synthetic testing visualizations.}}
	 \vspace{-3mm}
	\label{fig:box_three_datasets}
\end{figure}

\begin{figure}
    \centering
    \includegraphics[width=1\linewidth]{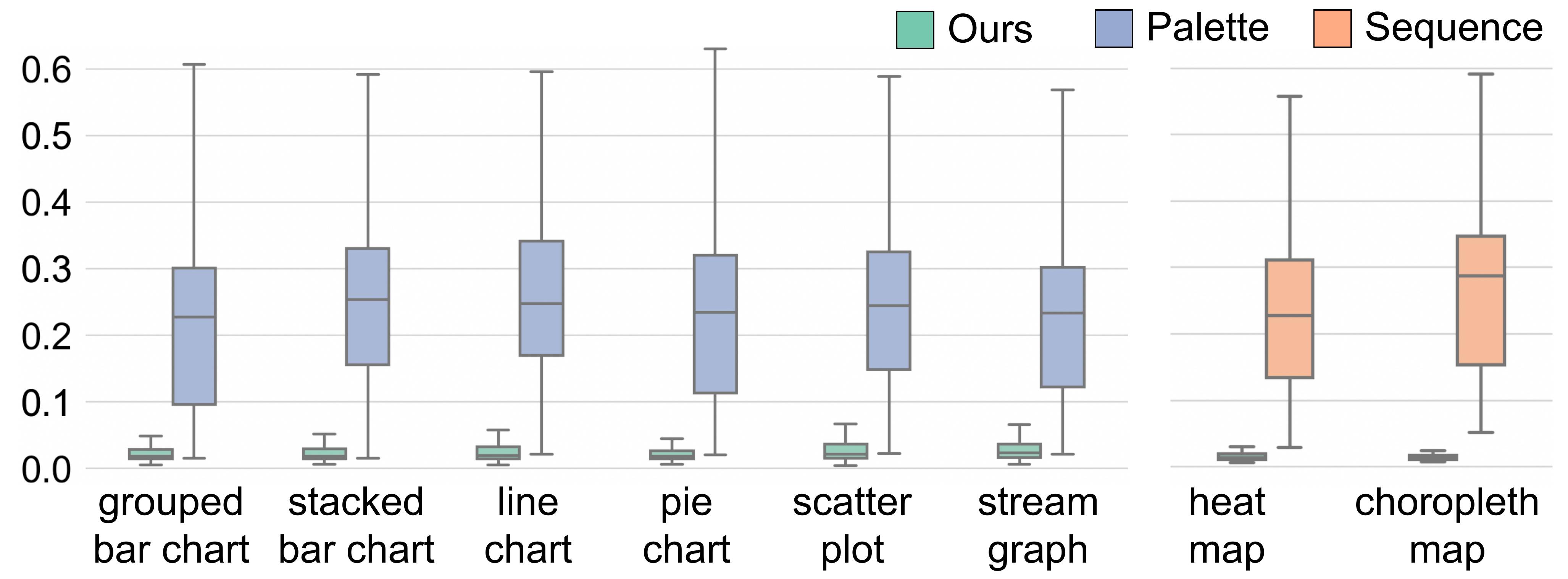}
    \vspace{-5mm}
    \caption{Comparing $D_{dtw}$ on the \tvcg{synthetic testing} visualizations of different chart types.
    Our method achieves better performances in all chart types, as indicated by the smaller $D_{dtw}$ values.
    }
     \vspace{-3mm}
    \label{fig:box_ctype}
\end{figure}

\begin{figure*}[t]
    \centering
    \includegraphics[width=0.995\linewidth]{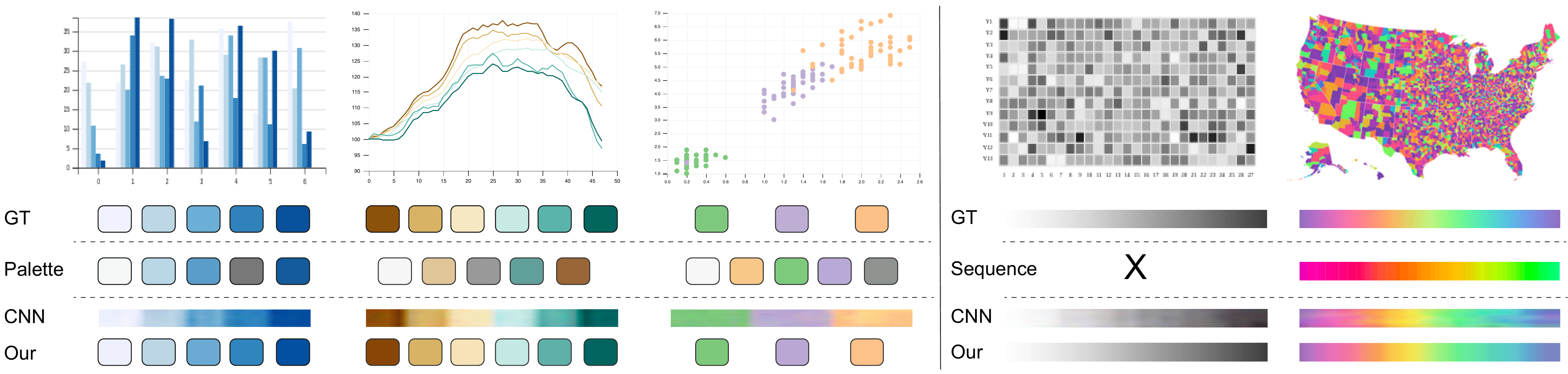}
    \vspace{-3mm}
    \caption{Example colormap results by our approach \emph{vs.} prior works. From top to bottom: input visualizations, ground-truth, colormaps extracted by palette-based~\cite{chang_2015_palette} or sequence-preserving~\cite{yoo_2015_color}, our CNN predictions, and the final results produced by our method.}
    \label{fig:comparison_synthetic}
     \vspace{-4mm}
\end{figure*}

\vspace{1mm}
\noindent
\textbf{Results}.
\tvcg{Fig.~\ref{fig:box_three_datasets} presents the normalized $D_{dtw}$ measured on three categories of synthetic testing, real seen, and real unseen visualizations. 
We first analyze the performance of our method across the three categories, and note that $D_{dtw}$ varies.
Our method achieves the best performance on synthetic testing visualizations, for both discrete (left) and continuous (right) colormaps. 
For real seen visualizations, the $D_{dtw}$ values increase, indicating the quality of predicted colormaps reduces.
There can be various reasons behind this, such as opaque or semi-transparent texts, low resolutions, and noise introduced to their color histograms.
The performance is better than that of real unseen visualizations, which have the highest $D_{dtw}$ values.
Since our approach is data-driven and the colormaps are unseen, it is not surprising that the performance drops.

Next, we compare our method with two baseline techniques across the three visualization categories. We conduct the independent \emph{t}-tests to examine the statistical significance among different methods. Specifically,

\begin{itemize}
    \item 
    \textit{Synthetic testing.}
    Our method significantly outperforms both palette-based ($t=-91.3, p<0.0001$) with mean $D_{dtw}$ reduced from \tvcgminor{$0.250_{[0.245,0.254]}$} (denoting mean with $95\%~CIs$) to $0.029_{[0.028,0.030]}$ (88\% improvement) for discrete colormaps, 
    and sequence-preserving  ($t=-50.1, p<0.0001$) method with mean $D_{dtw}$ reduced from \tvcgminor{$0.244_{[0.235,0.253]}$ to $0.015_{[0.014,0.016]}$} (94\% improvement) for continuous colormaps.

    \item
    \textit{Real seen.} 
    Our method still achieves significant smaller $D_{dtw}$ than both palette-based ($t=-5.1, p<0.0001$) and sequence-preserving  ($t=-8.1, p<0.0001$) method. 
    Yet, the improvements drop to 38\% (mean $D_{dtw}$ reduced from \tvcgminor{$0.240_{[0.212,0.269]}$ to $0.149_{[0.128,0.171]}$}) for discrete colormaps, and to 54\% (mean $D_{dtw}$ reduced from \tvcgminor{$0.227_{[0.192,0.263]}$ to $0.104_{[0.093,0.115]}$}) for continuous colormaps.
    
    \item
    \textit{Real unseen.}
    No significant difference is observed between ours and palette-based method ($t=-1.6, p=0.12$), whilst ours still outperforms sequence-preserving method with mean $D_{dtw}$ reduced from \tvcgminor{$0.332_{[0.283,0.380]}$ to $0.206_{[0.172,0.241]}$} (38\% improvement) for continuous colormaps.
\end{itemize}
}

Finally, we further compare three methods on synthetic testing visualizations in a finer-grained level by breaking down the results according to chart types, as shown in Fig.~\ref{fig:box_ctype}.
Overall, our approach outperforms both palette-based and sequence-preserving methods.
In more detail:

\begin{itemize}

\vspace{1mm}
\item
\textit{Discrete colormaps.}
Our method achieves significantly smaller $D_{dtw}$ than the palette-based method for all six chart types with discrete colormaps ($t=-91.3, p < 0.0001$).
More specifically, mean $D_{dtw}$ is reduced from \tvcgminor{$0.218_{[0.208,0.229]}$ to $0.026_{[0.024,0.028]}$} for \textit{grouped bar chart} (88.1\% improvement), 
from \tvcgminor{$0.263_{[0.251,0.275]}$ to $0.029_{[0.027,0.031]}$} for \textit{stacked bar chart} (89.0\% improvement), 
from \tvcgminor{$0.260_{[0.248,0.273]}$ to $0.029_{[0.027,0.031]}$} for \textit{line chart} (88.8\% improvement), 
from \tvcgminor{$0.222_{[0.212,0.232]}$ to $0.026_{[0.024,0.028]}$} for \textit{pie chart} (88.3\% improvement), 
from \tvcgminor{$0.259_{[0.248,0.270]}$ to $0.034_{[0.031,0.037]}$} for \textit{scatter plot} (86.9\% improvement), 
and from \tvcgminor{$0.243_{[0.231,0.253]}$ to $0.035_{[0.032,0.037]}$} for \textit{stream graph} (85.6\% improvement).

\vspace{1mm}
\item
\textit{Continuous colormaps.}
Our method also outperforms the sequence-preserving method for both \textit{heat map} and \textit{choropleth map} ($t=-50.0, p < 0.0001$).
Specifically, mean $D_{dtw}$ is reduced from \tvcgminor{$0.236_{[0.226,0.246]}$ to $0.016_{[0.015,0.017]}$} for \textit{heat map} (93.2\% improvement), and from \tvcgminor{$0.286_{[0.265,0.305]}$ to $0.014_{[0.013,0.015]}$} for \textit{choropleth map} (95.1\% improvement).

\end{itemize}

\tvcg{More quantitative comparison results can be found in Supplementary Sec. 4.1.}

\subsection{Qualitative Examples}
\label{ssec:example}
\textbf{Synthetic visualizations}.
Fig.~\ref{fig:comparison_synthetic} presents five visualization images from the evaluation dataset, each paired with the ground-truth colormap (GT), colormap by palette-based~\cite{chang_2015_palette} (Palette) or sequence-preserving~\cite{yoo_2015_color} (Sequence), our CNN prediction (CNN), and the final colormap produced by our method (Final).

The left side presents a bar chart, line chart, and scatterplot, respectively.
They are encoded with discrete colormaps shown in the GT row.
Palette~\cite{chang_2015_palette} shows two main deficiencies:
first, the method by default extracts five colors, whilst ground-truth colormaps usually have more variants, e.g., six in the line chart and three in the scatterplot.
Second, the method recovers color orders according to luminance channel, which often violates the color design rules.
Moreover, the method often extracts a gray color that is actually not in the colormap.
This is probably caused by the rendering process when drawing black grids and texts.

The fourth column presents a heatmap with grayscale colormap, which is not manageable by Sequence~\cite{yoo_2015_color}.
The method also produces incorrect colormaps for the US map in the last column, which is encoded with a rainbow colormap from the D3 library.
Here, we can see that purple colors are not available anymore, which may be projected to red colors that exhibit similar hues with purple;
and the blue colors are also omitted, probably because there are not many blue pixels in the map.

\begin{figure}[t]
    \centering
    \includegraphics[width=0.995\linewidth]{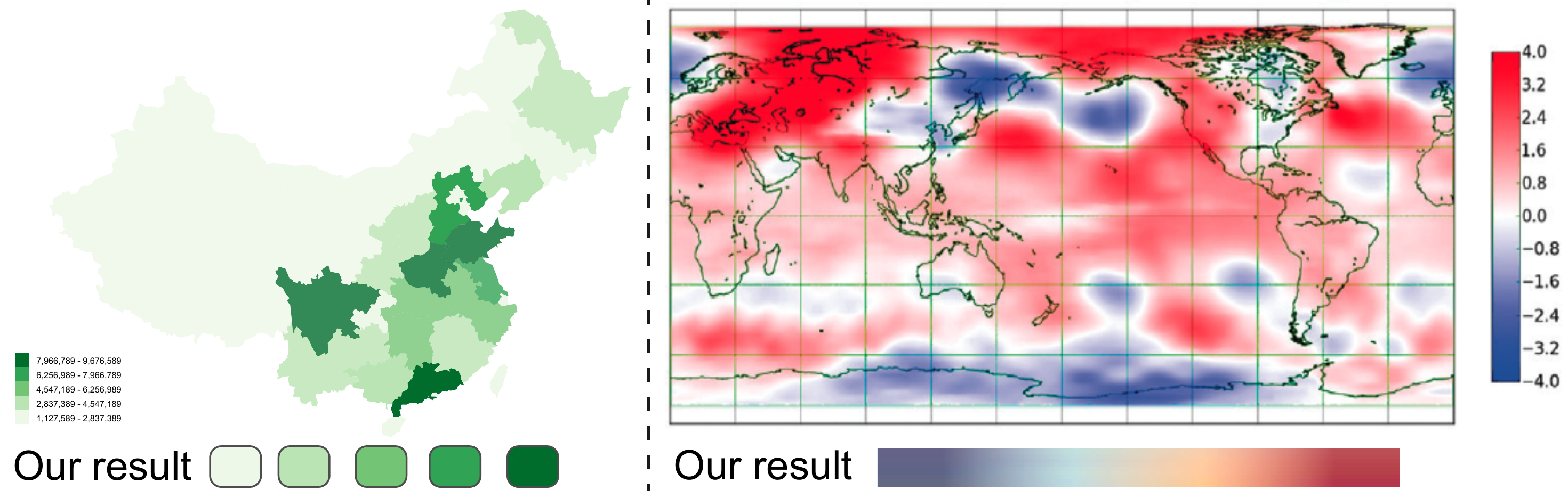}
    \vspace{-5mm}
    \caption{Unseen visualizations and results by our method.}
    \label{fig:comparison_real}  
     \vspace{-4mm}
\end{figure}

\begin{figure*}[t]
    \centering
    \includegraphics[width=0.995\linewidth]{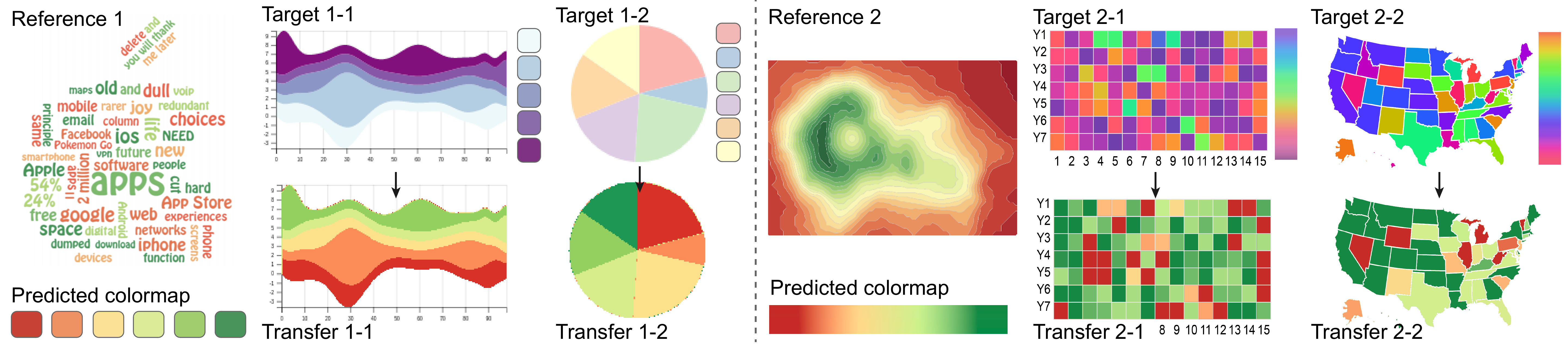}
    \vspace{-2mm}
    \caption{
    Transferring colormaps from reference to target visualizations.
    Left reference: a word cloud~\cite{wang_2018_edwordle} with a 6-class RdYlGn colormap.
    Right reference: a contour plot with a continuous RdYlGn colormap subject to the perceptual design principles.
    }
     \vspace{-4mm}
    \label{fig:recoloring_dis}
\end{figure*}

\vspace{2mm}
\noindent
\textbf{Real-world visualizations}.
Fig.~\ref{fig:comparison_real} presents two more results of real-world visualizations.
Corresponding colormap predictions by our method are presented underneath.
Note here the visualizations are unseen to our neural network model.
The left side presents a simple case of 5 single-hue discrete colors.
Both Palette~\cite{chang_2015_palette} method and ours can correctly extract the colormaps.
The right side presents a more complicated example from~\cite{poco_2018_extracting}.
Our method extracts a colormap that is very close to the ground truth.
Yet, we can notice some differences:
(i) The prediction does not include white color; this may be because the background filtering wrongly regards all white pixels as background;
and (ii) the prediction generates dark red, rather than the red color as in the ground truth.
Through a deep probe into the training data, we notice that the ground truth colormap is not in the corpus, and the network already tries its best to predict the most similar one.
More real-world examples can be found in Supplementary Fig. 9 and Fig. 10.

\subsection{Result Analysis}
\label{ssec:analysis}

\tvcg{The experiment results show a big difference between synthetic and real-world visualizations by our approach, whilst those by palette-based and sequence-preserving methods remain almost the same.
This is because our deep-learning-based approach is data-driven, of which the performance is affected by the difference between training and testing data distributions.
Our model is trained on synthetic visualizations of limited combinations of chart types, data distributions, and colormaps, whilst real-world visualizations feature new chart types, different data distributions, and unseen colormaps.
There is a dataset shift problem for real-world visualizations, yielding higher $D_{dtw}$ than synthetic visualizations.
On the other hand, palette-based and sequence-preserving methods are heuristic-based approaches, whose performances are highly dependent on hyperparameter tuning.
Nevertheless, it is challenging to find optimal hyperparameters for diverse visualizations, causing less accurate predictions than our approach.
}

We also notice that the performance of prior methods, especially Sequence~\cite{yoo_2015_color}, are unstable and error-prone. 
We have striven to achieve reliable results for them by carefully adjusting their parameter settings.
Nevertheless, our reimplementation may still bring in certain errors.
Without considering these errors, we can reasonably come to the following conclusions by probing deep into the quantitative results and representative examples:

\begin{itemize}
\item
\textit{Discrete colormaps.}
Palette~\cite{chang_2015_palette} produces high distance $D_F(C_{gt}, C_{out})$ mainly for two reasons:
(i) the method employs a fixed $k = 5$ for $k$-means clustering algorithm, thus it always generates five colors no matter how many colors are included in ground truth colormaps, as revealed by the line chart and scatterplot in Fig.~\ref{fig:comparison_synthetic}.
(ii) After retrieving the colors, the method sorts all the colors according to their luminance.
This simple heuristic could easily generate wrong color orders, as in Fig.~\ref{fig:comparison_synthetic}.

\vspace{1mm}
\item
\textit{Continuous colormaps.}
Sequence~\cite{yoo_2015_color} often wrongly group colors with similar hues, or omits colors with low histograms.
These heuristics can greatly affect the performance, in case of non-uniformly-distributed data; see the last example in Fig.~\ref{fig:comparison_synthetic} for example.
Thanks again to the CNN model, our method is more robust to these situations.
\end{itemize}

Our method can effectively address these problems by processing fine predictions generated by the CNN, instead of directly processing visualization images.
From Fig.~\ref{fig:comparison_synthetic}, we can observe that the intermediate predictions exhibit these properties:
(i) clear boundaries between distinct colors for discrete colormaps, promoting easy adaption of DBSCAN clustering instead of fix-sized $k$-means; 
(ii) well-balanced color histograms for continuous colormaps, making it easier for performing dimension reduction~\cite{yoo_2015_color}; and (iii) roughly ordered colors, which can be used as reference for recovering the color orders.  

\begin{figure*}[t]
  \centering 
  \includegraphics[width=0.95\textwidth]{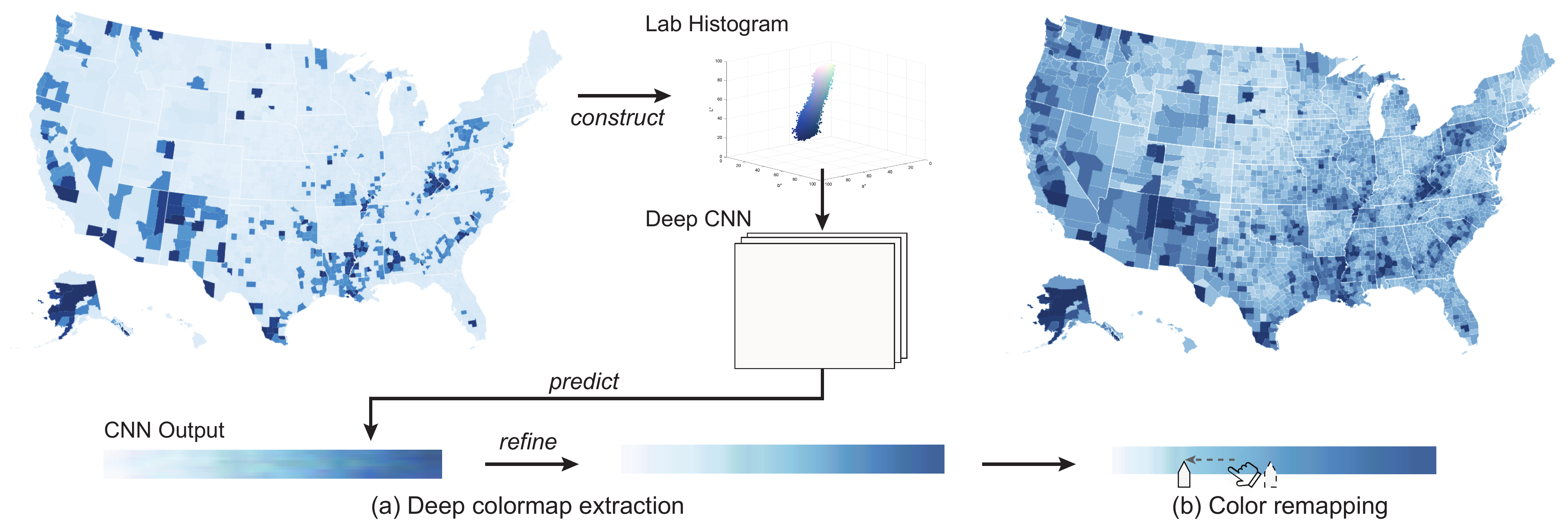}
  \vspace{-3mm}
      \caption{From the input visualization image (without a given colormap) shown on top left, 
(a) we first construct a color histogram in Lab color space, then employ our trained deep convolutional neural network (CNN) to predict a colormap and refine the colormap; by then, (b) we can remap the predicted colormap to refine the visualization.}
\label{fig:teaser}\vspace{-3mm}
\end{figure*}

\section{Applications}
\label{sec:app}

This section introduces two applications based on colormaps produced by our methods: 
color design transfer (Sec.~\ref{sec:example_1}) and color-coding adaptation (Sec.~\ref{sec:example_2}).

\subsection{Color Design Transfer}
\label{sec:example_1}

Inspired by example-based design, we can transfer color design from an existing visualization, by applying the extracted colormap to another visualization.
Here, we take two visualization images as input, one as \textit{reference} and the other as \textit{target}.
Our method automatically extracts two sorted lists of colors, denoted as $C_{ref} := \{c_{ref}^i\}^m_{i=1}$ and $C_{tgt} := \{c_{tgt}^i\}^n_{i=1}$, respectively.
Specifically, we constrain the number of colors in the target visualization to be less than that in the reference visualization, i.e., $n \leq m$.
Upon these conditions, we simply transfer the first $n$ colors from $C_{ref}$ to the target visualization.
To accomplish this, we replace pixels in the target visualization of color $c_{tgt}^j$ to color $c_{ref}^j$, where $1 \leq j \leq n$.

Fig.~\ref{fig:recoloring_dis} shows some example results.
The left one takes word cloud from a recent InfoVis paper~\cite{wang_2018_edwordle} as the reference.
It employs a conventional 6-class RdYlGn colormap from ColorBrewer and creates a pleasant visual appearance.
Our method successfully extracts the colormap presented underneath.
We transfer this colormap to (i) a ThemeRiver visualization with a 5-class BuPu colormap, which is difficult to distinguish light blue with white background;
and (ii) a pie chart with a 6-class Pastel1 colormap, which also includes a light yellow color that is close to the white background.
On the right side, a colored contour plot uses a continuous RdYlGn colormap.
Our method again successfully extracts its colormap presented underneath.
We then apply the colormap to refine a heatmap and choropleth map using rainbow colormaps that are not subject to human perception.
After imitation, we can more easily compare the data values in the refined visualizations.

\subsection{Color Remapping}
\label{sec:example_2}

Even with a well-chosen colormap, visually unappealing visualizations can still be created when inappropriate color ranges are applied to encode the unevenly-distributed data.
In such scenarios, it is hard to perceive correct data values from a visualization.
For instance, when a skewed dataset follows exponential or logarithmic distribution while a linear colormap is employed, only a small range of colors will be shown in the visualization.
Tominski et al.~\cite{tominski_2008_task} proposed to remap the color-coding in alignment with data distribution.
Inspired by them, we develop a customized interface that allows for color remapping.

This application is designed for continuous colormaps.
Specifically, we implement an intuitive \textit{median-color} slider beneath the extracted colormap; see Fig.~\ref{fig:teaser}(b).
Users can drag the slider to left/right to adjust the color-coding.
An algorithm is developed to automatically resample the colors along the sequential color graph~\cite{yoo_2015_color} based on the adjusted slider position.
Fig.~\ref{fig:teaser} illustrates a working example of the application.
Here, a linear \textit{BrBG} continuous colormap is employed to encode exponentially-distributed demographic data.
Inappropriate mapping between data values and color ranges makes the choropleth map filled with brown colors.
By dragging the \textit{median-color} slider towards left, a better visualization (Fig.~\ref{fig:teaser}(b)) is generated.
\section{Discussion}
\label{sec:discussion}

\begin{figure}[t]
	\centering
	\includegraphics[width=0.985\linewidth]{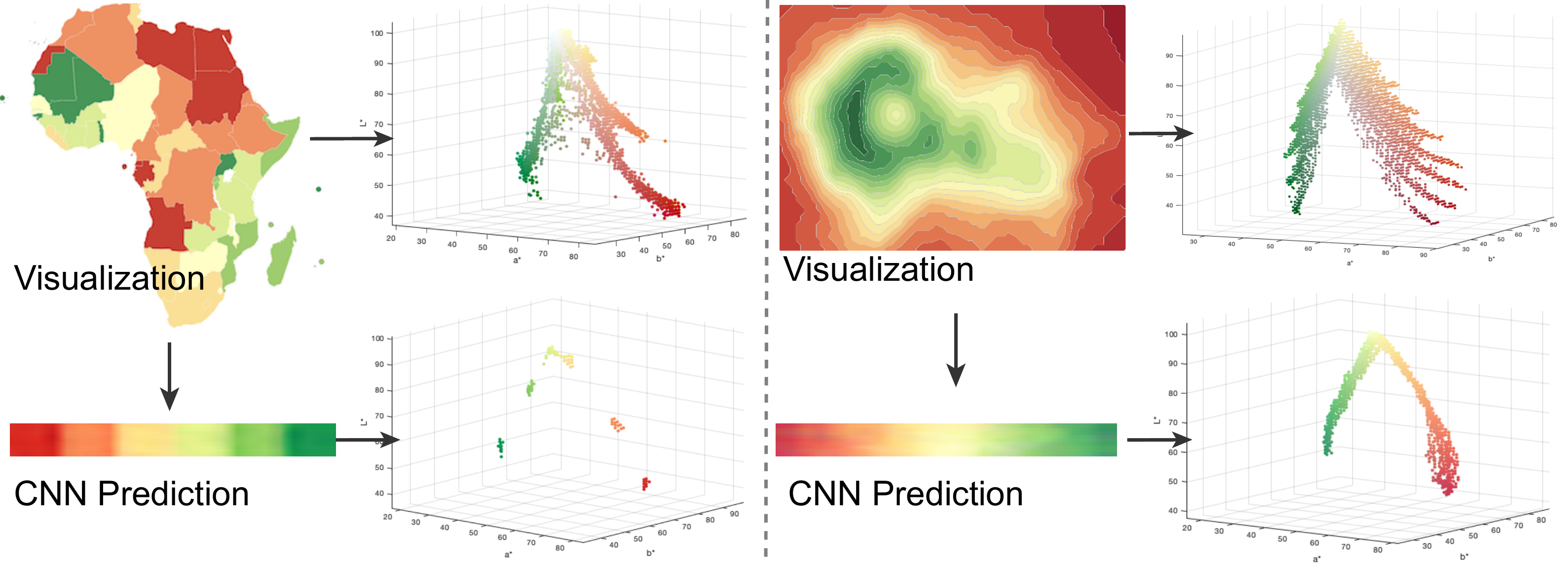}
	\vspace{-2mm}
	\caption{Color histograms constructed from the CNN predictions exhibit much distinctive difference between continuous and discrete colormaps than from the input visualizations.}
	\label{fig:why_cnn}
	 \vspace{-4mm}
\end{figure}

\subsection{Deep Learning vs. Heuristic Approaches}

The basis of our approach is a deep CNN model that learns to predict colormaps from input visualizations.
The prediction can be regarded as a mapping process from highly complex 3D histograms of original visualizations to relatively simple 2D colormaps.
By this, our approach bypasses cumbersome parameter settings adopted in prior heuristic approaches, \emph{e.g.}, thinning parameters in~\cite{yoo_2015_color}, and color distance threshold in~\cite{chang_2015_palette}.
In comparison with input visualizations, the predicted colormaps exhibit several distinctive benefits.
First, the predictions provide much evident distinction between continuous and discrete colormaps.
Fig.~\ref{fig:why_cnn} presents a comparison of color histograms constructed directly from input visualizations and from corresponding predictions.
Both visualizations employ RdYlGn colormaps: discrete for the left, and continuous for the right.
However, color histograms constructed directly from the visualizations look similar to each other.
In comparison, we can easily observe the differences between color histograms constructed from the predictions.
Second, recovering color ordering is crucial for visualization, yet the task is very challenging.
Many prior works employ simple heuristics based on hand-crafting the features to sort the colors, e.g., luminance values~\cite{chang_2015_palette}.
This can easily cause inappropriate color ordering; see Fig.~\ref{fig:comparison_synthetic} for examples.
On the other hand, CNNs learn the features to make the predictions, which naturally contain the color ordering information.

\subsection{Lessons Learned}
We observe that the CNN model produces more accurate predictions for colormaps with a small number of colors (less than seven), and with more distinctive colors (e.g., multiple-hue categorical colormaps); see Supplementary Fig. 8 for details.
Taking Fig.~\ref{fig:comparison_synthetic} as an example, CNN predictions for the leftmost bar chart (with five-class single-hue colormap) is relatively less accurate than the second line chart (with six-class diverging colormap) and the third scatterplot (with three-class multi-hue colormap).
Nevertheless, single-hue discrete colormaps with too many colors are not encouraged, as they can impose difficulty for humans to perceive.

Although the network is trained on synthetic visualizations of fixed resolution 512$\times$256, the CNN model is not constrained by the input image size.
This is because our CNN model consumes a Lab color histogram as input, instead of directly consuming the input visualization image.
Actually, our approach is not constrained to chart types and image sources as well.
For instance, both the word cloud and contour plot used in Sec.~\ref{sec:example_1} are unseen by the CNN model, and yet, our method can still extract their colormaps.

Besides the model trained with Lab color histograms, we also experimented with CNN models trained with the original visualization images and HSV color histograms as inputs.
As depicted by Fig.~\ref{fig:rotation}, the image model can not preserve geometric invariance, which is regarded as a core requirement for deep neural networks for visualization~\cite{haehn_2019_evaluating}.
In contrast, both the Lab and HSV models fulfill the requirement.
On the other hand, Lab and HSV models achieve very similar training performances; see Supplementary Fig. 4 for details.
We envision that a combination of both color spaces can enrich the feature map for a CNN to learn, which may lead to even better colormap prediction, as in the case of image classification~\cite{DBLP:journals/corr/abs-1902-00267}.

\subsection{Limitations}
Though comparably effective and robust, our approach also exhibits several limitations.

First, our approach is data-driven, meaning that the network learns to predict based on the training dataset.
This means that if there are new input patterns that the network has not seen before, it may produce poor results.
Hence, we have striven to synthesize a dataset that covers a wide variety of visualizations in different styles by carefully considering data distributions, chart types, and colormaps.
Yet, the space for visual design is too vast to be entirely covered.
For instance, our synthetic dataset does not include tree-structured colors~\cite{tennekes_2014_tree} and semantically meaningful colormaps~\cite{lin_2013_selecting}.

Second, our approach works on color histograms rather than original images, limiting the scope of our work to linear colormaps only.
Nevertheless, this is a common limitation for all existing colormap extraction methods that work on a color space~\cite{yoo_2015_color}, as no algorithm can explicitly distinguish if the histogram imbalance is caused by the underlying data distribution or by a non-linear colormap.
We design an interactive interface that allows users to create nonlinear colormaps (Sec.~\ref{sec:example_2}).

Third, colormaps retrieved by our approach only indicate relative color ordering without explicit data semantics.
That is, we can only tell if a color is ahead of or behind another one, but not able to infer what data value it represents.
This limitation constrains our approach in applications based on both data and color mapping, such as interactive overlay~\cite{poco_2018_extracting}.
Again, this is a common limitation also for existing colormap extraction methods.
A complementary solution is to detect and recognize text labels in original images, and associate them with the retrieved colormaps~\cite{poco_2018_extracting}.
\section{Conclusion and Future Work}
\label{sec:conclusion}

In this paper, we have presented a new approach to automatically extract colormaps from visualizations.
The core of our method is a deep neural network that learns the mapping from Lab histograms to colormaps.
To promote effective learning, we have 
i) synthesized a new dataset of diverse visualizations that covers a wide range of data distributions, chart types, and colormaps; 
ii) employed several feasible adaptions to the neural network, including conversion of 3D Lab histograms to 2D maps to fit with the CNN architectures, ASPP modules to create more nonlocal features, and fixed-size output image to handle conflict of limited neurons \emph{vs.} enormous colors.
Comparisons with state-of-the-art colormap extraction methods, sequence-preserving~\cite{yoo_2015_color} and palette-based~\cite{chang_2015_palette}, also confirmed the advantages of our method; see both the quantitative evaluations and several representative examples.
In the end, we presented two applications that can benefit from our method, color design transfer and color remapping.

Our work opens several directions for future research.
First, our approach allows fully-automatic extraction of colormaps, enabling us to feasibly process vast amounts of visualization images.
Inspired by TreeVis.net~\cite{schulz_2011_treevis} and VIStory~\cite{zeng_2021_vistory}, we plan to exploit publications in visualization conferences and journals, and create a reference website that summarizes color usage by the community.
Second, a great challenge encountered in this work is the lack of proper training data.
We will release the dataset to benefit future researches.
Last, our network learns distinct features among color histograms by consuming vast amount of visualizations and corresponding colormaps.
\tvcg{We would like to further enrich the training dataset by considering tree-structured colormaps~\cite{tennekes_2014_tree}, semantic colormaps~\cite{lin_2013_selecting}, and user-created continuous colormaps~\cite{nardini_2019_making}, and add some noise to the synthetic visualizations to better cope with real-world visualizations.}
Nevertheless, a more appropriate approach would be to directly learn color design \textit{rules}.
\tvcgminor{It would be interesting to try using a graph neural network~\cite{wu_2019_comprehensive}.}
\section*{Acknowledgment}
The authors wish to thank the anonymous reviewers for their valuable comments.
This work is supported in part by National Natural Science Foundation of China (61802388) and SIAT Innovation Program for Excellent Young Researchers.
\bibliographystyle{abbrv}
\bibliography{Reference}

\begin{IEEEbiography}[{\includegraphics[width=1in,height=1.25in,clip,keepaspectratio]{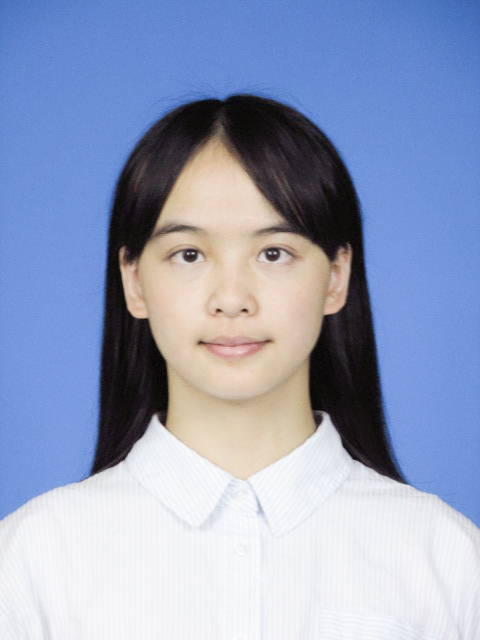}}]
{Lin-Ping Yuan} is currently a Ph.D. student in the Department of Computer Science and Engineering at the Hong Kong University of Science and Technology (HKUST). She obtained her B.Eng. degree in Software Engineering from Xi’an Jiaotong University, China in 2019. Her research interests include information visualization, visual analytics, and image processing.
\end{IEEEbiography}

\begin{IEEEbiography}[{\includegraphics[width=1in,height=1.25in,clip,keepaspectratio]{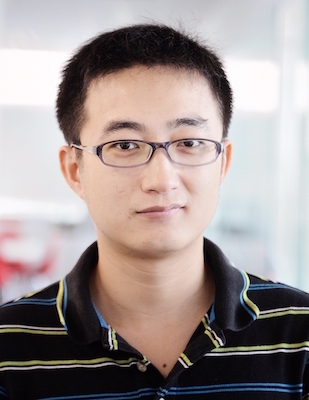}}]
{Wei Zeng} is currently an associate professor at Shenzhen Institutes of Advanced Technology, Chinese Academy of Sciences.
He received his bachelor's degree (2011) and Ph.D. (2015), both in computer science, from Nanyang Technological University.
He served as program committee members in various research conferences, including IEEE VIS, ChinaVis, PacificVis Poster, IVAPP.
His research interests include all aspects of data visualization, with a particular focus on interactive techniques for urban data analysis, and data augmented design.
\end{IEEEbiography}

\begin{IEEEbiography}[{\includegraphics[width=1in,height=1.25in,clip,keepaspectratio]{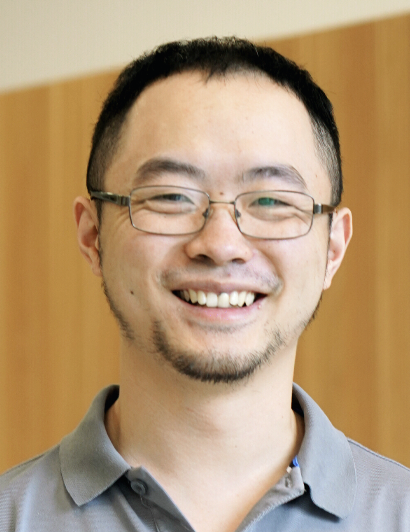}}]
{Siwei Fu} is an associate research scientist in Zhejiang Lab. He received his Ph.D. degree in the Department of Computer Science and Engineering at the Hong Kong University of Science and Technology. His main research interests are in visualization and human-computer interaction, with focuses on visual text analytics, multidimensional data visualization, and visualization recommendation.
\end{IEEEbiography}

\begin{IEEEbiography}[{\includegraphics[width=1in,height=1.25in,clip,keepaspectratio]{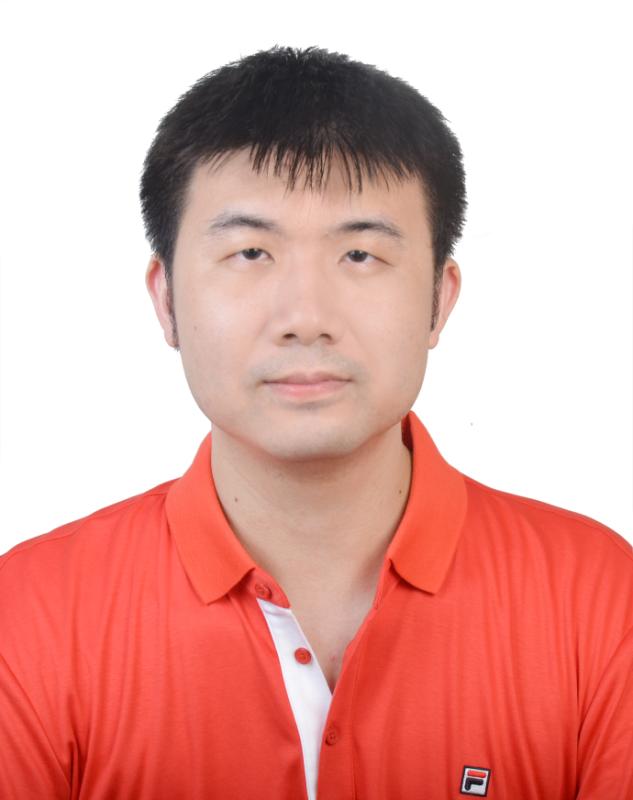}}]
{Zhiliang Zeng} received the B.S. degree from the Beijing Normal University, Zhuhai, and the M.Sc. degree in Computer Science and Engineering from the Chinese University of Hong Kong. 
He is currently a PhD student in Computer Science and Engineering of Chinese University of Hong Kong.
His research interests include deep neural network and image segmentation, using neural network for indoor/outdoor scene analysis and application. 
\end{IEEEbiography}

\begin{IEEEbiography}[{\includegraphics[width=1.0in,height=1.25in,clip,keepaspectratio]{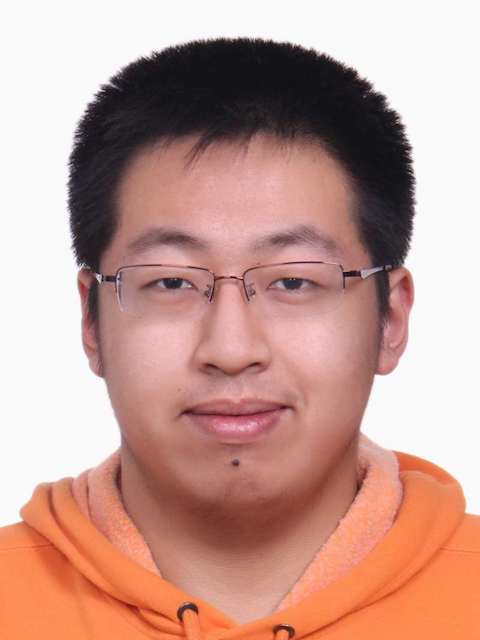}}]{Haotian Li}
Haotian Li is currently a PhD student in Computer Science and Engineering at the Hong Kong University of Science and Technology (HKUST). His main research interests are data visualization, visual analytics and data mining, with emphasis on online learning and fintech. He received his BEng in Computer Engineering from HKUST. For more details, please refer to https://haotian-li.com/.
\end{IEEEbiography}

\begin{IEEEbiography}[{\includegraphics[width=1.0in,height=1.25in,clip,keepaspectratio]{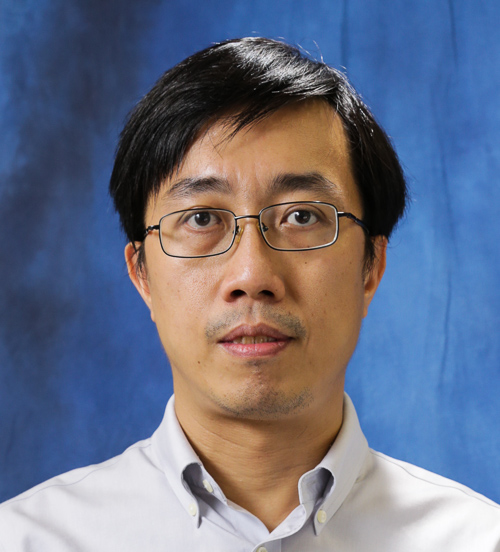}}]{Chi-Wing Fu} is currently an associate professor in the Chinese University of Hong Kong.  He served as the co-chair of SIGGRAPH ASIA 2016's Technical Brief and Poster program, associate editor of IEEE Computer Graphics \& Applications and Computer Graphics Forum, panel member in SIGGRAPH 2019 Doctoral Consortium, and program committee members in various research conferences, including SIGGRAPH Asia Technical Brief, SIGGRAPH Asia Emerging tech., IEEE visualization, CVPR, IEEE VR, VRST, Pacific Graphics, GMP, etc.  His recent research interests include computation fabrication, point cloud processing, 3D computer vision, user interaction, and data visualization.
\end{IEEEbiography}

\begin{IEEEbiography}[{\includegraphics[width=1.0in,height=1.25in,clip,keepaspectratio]{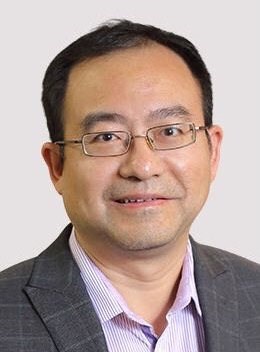}}]{Huamin Qu} is a professor in the Department of Computer Science and Engineering (CSE) at the Hong Kong University of Science and Technology (HKUST) and also the director of the interdisciplinary program office (IPO) of HKUST. He obtained a BS in Mathematics from Xi'an Jiaotong University, China, an MS and a PhD in Computer Science from the Stony Brook University. His main research interests are in visualization and human-computer interaction, with focuses on urban informatics, social network analysis, E-learning, text visualization, and explainable artificial intelligence (XAI).
\end{IEEEbiography}

\vfill


\end{document}